\documentclass[preprint,aps,prc,amsmath,amssymb,showpacs,showkeys,floatfix,nofootinbib]{revtex4-2}

\usepackage{dcolumn}
\usepackage{bm}
\usepackage{array,tabularx}
\usepackage{multirow}
\usepackage{xcolor}
\usepackage{amsthm}
\usepackage{amssymb}
\usepackage[flushleft]{threeparttable}
\usepackage{enumerate}

\newcommand{\braket}[3]{\bigl\langle#1\bigr|#2\bigl|#3\bigr\rangle}
\newcommand{\Braket}[2]{\bigl\langle#1|#2\bigr\rangle}

\usepackage[pdftex]{graphicx}
\usepackage[pdftex]{hyperref}
\hypersetup{bookmarksnumbered=true,colorlinks=true,linkcolor=magenta,citecolor=blue,breaklinks=false}

\usepackage[a4paper,left=0.5cm,right=0.5cm,top=1cm,bottom=1cm,includefoot=true,
ignorehead=false,ignoremp=true,driver=dvips,footskip=2cm,headheight=0cm,headsep=0cm]{geometry}

\providecommand{\R}{\ensuremath{\mathbb{R}}}
\providecommand{\M}{m_{\text{th}}}
\providecommand{\al}{\alpha}
\providecommand{\ga}{\gamma}
\providecommand{\Ga}{\Gamma}
\providecommand{\om}{\omega}
\providecommand{\lam}{\lambda}

\providecommand{\II}{\ensuremath{\mathcal{I}}}
\newcommand{\EL}[1]{\Biggl\{#1\Biggr\}} 
\newcommand{\GL}[1]{\biggl\{#1\biggr\}} 
\newcommand{\NL}[1]{\bigl\{#1\bigr\}}   

\newcommand{\EC}[1]{\Biggl[#1\Biggr]}   
\newcommand{\GC}[1]{\biggl[#1\biggr]}   
\newcommand{\MC}[1]{\Bigl[#1\Bigr]}     
\newcommand{\NC}[1]{\bigl[#1\bigr]}     

\newcommand{\EP}[1]{\Biggl(#1\Biggr)}   
\newcommand{\GP}[1]{\biggl(#1\biggr)}   
\newcommand{\MP}[1]{\Bigl(#1\Bigr)}     
\newcommand{\NP}[1]{\bigl(#1\bigr)}     

\newcommand{\NB}[1]{\bigl|#1\bigr|}     
\newcommand{\MB}[1]{\Bigl|#1\Bigr|}     
\newcommand{\GB}[1]{\biggl|#1\biggr|}   
\newcommand{\EB}[1]{\Biggl|#1\Biggr|}   

\renewcommand{\Re}{\text{Re\,}} 
\renewcommand{\Im}{\text{Im\,}} 
\DeclareMathOperator{\Arg}{Arg} 
\DeclareMathOperator{\Res}{Res}

\providecommand{\dd}{\tilde{\delta}}
\begin{document}

\title{Quantum Corrections to the Decay Law in Flight}
\author{D. F. Ram\'irez Jim\'enez}
\email{df.ramirez@doctoral.uj.edu.pl}
\affiliation{M. Smoluchowski Institute of Physics,
Faculty of Physics, Astronomy and Applied Computer Science,
Jagiellonian University,  PL-30348 Krak\'ow, Poland}
\author{A. F. Guerrero Parra}
\email{af.guerrerop@uniandes.edu.co}
\affiliation{Departamento de F\'isica, Universidad de los Andes,
Cra. 1E No. 18A-10, Bogot\'a, Colombia}

\author{N. G. Kelkar}
\email{nkelkar@uniandes.edu.co}
\affiliation{Departamento de F\'isica, Universidad de los Andes,
Cra. 1E No. 18A-10, Bogot\'a, Colombia}

\author{M. Nowakowski}
\email{marek.nowakowski@ictp-saifr.org}
\affiliation{ICTP South American Institute for Fundamental 
Research (ICTP-SAIFR)
IFT-UNESP, Rua Bento Teobaldo Ferraz 271, Bloco 2,
01140-070 S\~ao Paulo, SP, Brazil
}
\affiliation{Universidade Federal de S\~ao Paulo, C.P. 01302-907, S\~ao Paulo, 
Brazil}
\date{\today}

\begin{abstract}
The deviation of the decay law from the exponential is a well known effect of 
quantum mechanics. Here we analyze the relativistic survival probabilities, 
$S(t,p)$, where $p$ is the momentum of the decaying particle 
and provide analytical expressions for $S(t,p)$ in the exponential (E) as 
well as the nonexponential (NE) regions at small and large times. 
Under minimal assumptions on 
the spectral density function, analytical expressions for the 
critical times of transition from the NE to the E at small times and the 
E to NE at large times are derived. 
The dependence of the decay law on the relativistic Lorentz factor, 
$\gamma = 1/\sqrt{1 - v^2/c^2}$, reveals 
several interesting features. In the short time regime of the decay law, 
the critical time, $\tau_{st}$, shows a steady increase with $\gamma$, thus 
implying a larger NE region for particles decaying in flight. Comparing 
$S(t,p)$ with the well known time dilation formula, 
$e^{-\Gamma t/\gamma}$, in the exponential region, an expression for the 
critical $\gamma$ where $S(t,p)$ deviates most from $e^{-\Gamma t/\gamma}$ is
presented. This is a purely quantum correction. Under particular conditions 
on the resonance parameters, there also exists a critical $\gamma$ at large 
times which decides if the NE region shifts backward or forward in time as 
compared to that for a particle at rest. 
All the above analytical results are supported by calculations involving 
realistic decays of hadrons and leptons.  

\end{abstract}
\maketitle

\section{Introduction}
In the last two decades there have been new interesting developments,
both from the experimental as well the theoretical side, in the theory
of quantum mechanical time evolution of unstable states. Some anomalies
have been reported. One of them is regarding the time modulation of the 
exponential decay of $^{140}$Pr$^{59+}$ and $^{142}$Pm$^{60+}$ ions
which decay by orbital electron capture  
\cite{GSI}. Another one is related to the seasonal variation of 
nuclear beta decay rates \cite{sturrock}. 
A more recent experiment with $^{142}$Pm$^{60+}$ ions at GSI 
\cite{ozturk}, however, did not confirm the oscillations superimposed over 
the exponential decay. Explanations of the seasonal variation of the 
beta decay rates were sought through neutrinos emerging from the Sun and 
the Earth-to-Sun distance (see \cite{schrader} and references therein).   
The Sun's influence was ruled out in \cite{semkow} by finding an alternative 
explanation with temperature variations of the density of gas in the ionization
chamber. More exotic explanations such as a 
misalignment mechanism of QCD axion dark matter, based on an analysis of 
12 years of tritium decay data can be found in \cite{zhang}. 
Claims of proximity to the Sun causing variations in the decay constants 
have been investigated in \cite{pomme} using data measured over 5 decades 
at different nuclear institutes.  
These anomalies have spurned a 
number of theoretical papers \cite{GiraldiEPhysD,GiacosaQM,UrbanowskiAPPB} 
trying to explain them. In our
opinion, however, the most exciting theoretical advancement in the
area of metastable states
is the reconsideration of the exponential decay law in flight. 
Whereas there is no doubt about the exponential decay at
rest, a controversy has started  regarding it's form when the unstable
particle is moving while the observer is at rest. Special Theory of
Relativity predicts by time dilation, a simple substitution from
$\exp(-\Gamma t)$ to $\exp(-\Gamma t/\gamma)$ where $\Gamma \sim 1/\tau$ 
with $\tau$ being the mean
lifetime and $\gamma=1/\sqrt{1-v^2/c^2}$. This has been tested
and confirmed in various experiments (see the discussion in the next section)  
with a certain precision. Above all, there is a
famous experiment with atmospheric muons cited ubiquitously as a
paradigm of tests for Special Relativity \cite{muonexp1}. 
Starting with the paper by
Khalfin \cite{khalfinRel} who uses the Fock-Krylov method \cite{FockKrylov},
based on principles of quantum mechanics, a number of papers have
confirmed, rejected or modified the result in \cite{khalfinRel}, 
where the 
survival probability in flight was calculated as 
$\exp(-\sigma (p)t)$ with $\sigma (p)
\neq \Gamma /\gamma$. This form
depends on the momentum $p=m_0\gamma v$ and does not reduce to the
simple result of Special Relativity. It approximates it, however, for
narrow resonances for which $\Gamma=1/\tau\ll m_0$. Since we devote
the next full section on the status quo and history of the subject, we
will not dwell here on the details. It suffices to say that a
deviation from $\exp(-\Gamma t/\gamma)$ is not a contradiction to
Special Theory of Relativity. The exponential decay law at rest is
to a large extent a classical result and Special Relativity uses this
result to convert it into an expression valid in flight. Once quantum
mechanics is invoked, we might expect new effects in this domain.

Even if most of the realistic decay processes remain de facto
unaffected by the new corrections of \cite{khalfinRel}, it is, of course a
matter of principles to pin down the correct form of the exponential
decay in flight. At rest or otherwise, the latter appears in many
branches of physics where metastable states can be found. Therefore,
we could count the exponential decay as one of the important and
universal law of physics. For this reason we have decided to revisit
the subject with the emphasis on analytical expressions in the hope
that one day they might get tested experimentally and decide 
upon the controversy. In contrast to the
classical treatment of time evolution which encompasses only the
exponential form, quantum mechanics predicts a different behavior for
very small and very large times. In this context $\exp(-\Gamma t)$ is
only valid for intermediate times. This implies that one can also
look for a nonexponential behavior of the 
survival probability for small and large times when $p \neq 0$.
For small times we pick up the subject for the first time and
re-consider it also for large times. In both cases we obtain novel
results. We pay attention to the two transition times, from small times
to the intermediate and from intermediate to large and derive
expressions of the transition times in dependence of the parameters of
the decay. This is to say, in dependence of 
$\Gamma$, $m_0$, $m_{th}$ (the threshold mass) and now also $p$.

The article is organized as follows: in the next section, we present the 
different approaches in literature for the relativistic survival probability 
and how they compare with the standard time dilation formula. Choosing the 
most frequently used relativistic 
expression in literature, in Section (\ref{math_rel}) we analyze the 
mathematical properties of the survival amplitude $a(\tau, P)$ and 
probability $S(\tau,P)$ (with the dimensionless variables $\tau$ and $P$ being 
related to time and linear momentum respectively). This section 
contains the main results (both analytical and numerical) of the present work.  
Setting the grounds in Section (\ref{NRsurv}) with the known behaviours 
of the non-relativistic 
amplitude, $a(\tau, 0)$ and probability, $S(\tau,0)$ in the exponential 
and nonexponential regions, Section (\ref{intert}) presents the quantum 
corrections to exponential decay in flight. An analytical formula for the 
critical value of the Lorentz factor, $\gamma = 1/\sqrt{1 - v^2/c^2}$, is 
given along with numerical examples of some realistic resonances.   
A neat derivation of the the survival amplitude and probability in the large 
and short time regions is provided in Sections (\ref{larget}) and 
(\ref{shortt}) with detailed considerations of the threshold and the 
non-relativistic and ultra relativistic limits. Apart from the analytical 
formulae corresponding to the relativistic cases and their comparisons with 
the non-relativistic ones, some interesting features of the energy 
uncertainties corresponding to decays in flight are presented in this section.
In the last subsection of Section (\ref{math_rel}) we obtain expressions 
for the critical transition times from the short time non-exponential (NE) 
to the exponential (E) region and the (E) to large time (NE) region. Once 
again, interesting features with realistic resonances are noted. Finally, 
in Section (\ref{summ}) we summarize the results of the present work. 
 
\section{Survival probability of moving unstable systems}\label{section2}
The seminal work of Khalfin in 1957 \cite{khalfin1957} 
established that though the decay law of 
an unstable system can be shown classically to be of an exponential 
form, this description fails at short and large times. 
The former case is described by a quadratic function in $t$ 
\cite{levitan} and the latter by
an inverse power law in $t$ \cite{fondaghirardi}. 
Though theoretically established, the nonexponential behaviour of the 
decay law is hard to get experimentally \cite{weld, norman}.  
Indirect evidence of it's existence \cite{kelkar_nowakowski1} and the complete 
absence of the exponential decay for very broad resonances was however 
determined, based on scattering data \cite{kelkar_nowakowski2}.   
A few decades later, the 
survival probability of a particle moving with relativistic momenta became 
a topic of debate. The well known Einstein's time dilation formula predicts 
the increase in lifetime of a moving particle, $\tau_p$, as compared to that 
of a particle at rest, $\tau_0$, to be
\begin{equation}\label{lifetime}
\tau_p = \cfrac{\tau_0}{\sqrt{1 -\cfrac{v^2}{c^2}}} \,= \, \gamma \tau_0 \,,
\end{equation} 
where, $\NP{1 - v^2/c^2}^{-1/2}$ is the Lorentz factor denoted by $\gamma$. 
Defining the 
lifetime of the particle, in terms of the survival probability, $P(t)$, as, 
\begin{equation}\label{lifetime2}
\tau_0 = \int_0^{\infty} P_0(t) dt\, ,
\end{equation}
where, $P_0(t)$ is the survival probability of the particle at rest, it is 
easy to check that in case of a {\it purely exponential decay law}, 
\begin{equation}\label{specialrel}
P_p(t) = P_0(t/\gamma) \, ,
\end{equation} 
with $P_p(t)$ being the survival probability of the unstable 
particle in motion.
This result of relativistic time dilation has been verified 
experimentally \cite{firstexp} 
using muons \cite{muonexp1,muonexp2}, pions \cite{piexp}
and other relativistic particles \cite{other1,other2}.  
However, the validity of this result in a fully quantum mechanical 
calculation of $P_p(t)$ has been doubted in literature, thus 
providing corrections to the special relativity result in (\ref{specialrel}). 
In this section, different approaches leading to deviations from 
(\ref{specialrel}) will be briefly reviewed. Implications of these deviations 
for the survival probability at short and large times will be discussed for 
an isolated resonance by providing analytical expressions 
wherever possible.  
  
\subsection{Fock-Krylov method}\label{FK}
A standard approach for computing the survival amplitude 
is the Fock-Krylov (FK) method 
\cite{FockKrylov}. In order to understand the extension 
(or re-interpretation as we will see below) of the commonly 
used formula for computing the survival amplitude of moving unstable 
states, we briefly recapitulate the derivation of the standard formula 
presented in 
\cite{RamirezJPhysA}. The FK method involves expanding the initial state 
in eigenstates of a complete set of observables which commute with the 
Hamiltonian. 
Since the initial unstable state, $\vert \Psi (0) \rangle$, cannot be an
eigenstate of the (hermitian) Hamiltonian, an expansion in terms of energy 
eigenstates is used to write 
$\vert \Psi(0) \rangle$, at $t=0$ as,  
\begin{equation}\label{psi0}
\vert \Psi (0) \rangle =  \int dE \,\,a(E)\, \vert E \rangle\, .
\end{equation}
If we define the {\it survival amplitude $A^{FK}(t)$} as
\begin{equation}\label{def-surv-ampl}
A^{FK}(t)=\Braket{\Psi(0)}{\Psi(t)}=\braket{\Psi(0)}{e^{-iHt}}{\Psi(0)},
\end{equation}
we get the following by substituting $\vert \Psi(0) \rangle$ in the former definition:
\begin{align}\label{survival1}
A^{FK}(t) &= \int dE'\, dE \,\,a^*(E')\, a(E)\, \langle E'
\vert e^{(-iHt)} \vert E \rangle \notag\\
&= \int dE' \,dE \,\,a^*(E') \,a(E)\, e^{(-iEt)} \,\delta(E - E') \notag \\
&= \int \,dE\, |a(E)|^2\,e^{-iEt} = \int_{E_{th}}^{\infty} 
\,dE\, \omega(E) \, e^{-iEt}.
\end{align}
Here, $\omega(E)$ is the energy distribution or the so-called density of 
states of the unstable particle and this spectrum is bounded from below. 
$A^{FK}(t)$ 
is the amplitude of the probability that at the time $t$, the unstable 
particle will be in the initial undecayed state. 

In an attempt to combine quantum theory with relativity, Khalfin 
\cite{khalfinRel} noted that the density 
$\omega(E)$ appearing in the FK method, can actually be considered as 
a ``conditional density of the energy distribution of the unstable particle 
with momentum $p$" and hence, rewriting $A^{FK}(t)$ as 
\begin{equation}\label{survival2}
A_p^{Kh}(t) = \int_{SpecH} \,dE\, \omega(E|p) \, e^{-iEt}
\end{equation}
and using the relativistic energy-momentum relation, $E^2 = p^2 + m^2$, 
claimed that
\begin{equation}\label{khalfindensity}
\omega(E|p) dE = \bar{\omega} (m|p) dm \, ,
\end{equation}  
where, $\bar{\omega} (m|k)$, is a ``conditional density of the mass 
distribution of the unstable particle with momentum $p$". However, the 
mass distribution of an elementary unstable particle is invariant and 
cannot depend on the momentum. Hence, concluding that 
$\bar{\omega}(m|p) = \omega (m) $, he obtained, 
\begin{equation}\label{survival2p}
A_p^{FK}(t) = \int_{m_{\text{th}}} \,dm\, \omega(m) \, e^{-i\sqrt{p^2 + m^2}t}\, .
\end{equation}

Using a Breit-Wigner mass distribution and with a survival amplitude as 
in (\ref{survival2p}), 
Shirokov \cite{shirokov2004,shirokov2009,shirokov2009concepts} provided 
analytical formulae for the survival probability 
for large times. 

The same expression as in Eq. (\ref{survival2p}) was obtained by 
Urbanowski \cite{urbanowski_rel}  
by explicitly considering the transformation of the initial state from the 
rest frame of the decaying particle to the moving one.  
Rewriting Eq. (\ref{psi0}) for the case with zero momentum and denoting it 
as $\vert \Psi_0 (0) \rangle$, 
\begin{equation}\label{psi0_2}
\vert \Psi_0 (0) \rangle =  \int dm \,\,c(m)\, \vert m;0 \rangle\, .
\end{equation}
The author further notes that if $\Lambda$ denotes the Lorentz 
transformation then using  
$U(\Lambda)$ which is a unitary representation of the transformation, leads 
us to  
\begin{equation}\label{psip}
\vert \Psi_p(0) \rangle = U(\Lambda) \vert \Psi_0(0) \rangle = 
\int dm \,\,c(m)\,U(\Lambda) \vert m;0 \rangle \, .
\end{equation}
Now using, $\vert \Psi_p(0) \rangle$ (with the explicit momentum dependence) 
in order to define the survival amplitude $A_p^U(t)$ as
\begin{equation}
A_p^U(t)=\Braket{\Psi_p(0)}{\Psi_p(t)}=\braket{\Psi_p(0)}{e^{-iHt}}{\Psi_p(0)}
\end{equation} 
and repeating similar steps as in the Fock-Krylov method of Eq. (\ref{survival1}), 
one obtains,
\begin{equation}\label{survival_p}
A_p^U(t) \, = \, \int \,dm \, dm^{\prime} \, c^*(m^{\prime})\, c(m) \, 
 \langle m^{\prime}, 0 \vert U^{\dagger} e^{-iHt} U \vert m, 0 \rangle \, ,
\end{equation}
where, further noting that the operators $H, \vec{p}$ form a 4-vector $P_{\nu} = (P_0, \vec{p}) \equiv (H, \vec{p})$, where \cite{gibson}
\begin{align}   
U^{\dagger}(\Lambda) e^{-iHt} U(\Lambda) &= 
e^{-iU^{\dagger}(\Lambda)\,H \,U(\Lambda)},\\
U^{\dagger}(\Lambda)\,P_{\mu} \,U(\Lambda) &= \Lambda _{\mu}^{\,\,\,\nu} \, 
P_{\nu},
\end{align}
the author \cite{urbanowski_rel} finally obtains, 
\begin{equation}\label{survival3}
A_p^{U}(t) =
\int_{m_{th}}^{\infty} \,dm\, \omega(m) \, e^{-i\sqrt{p^2 + m^2}t}
\equiv A_p^{FK}(t)\, ,   
\end{equation}
where, $\omega(m) \equiv |c(m)|^2$. This expression, based on the principles 
of quantum mechanics, leads to the decay law in flight and is essentially 
different from that in (\ref{specialrel}).  
Noting that 
$E = m \gamma$, one could rewrite the exponential in the above equation 
as $e^{-i\sqrt{p^2 + m^2}t} = e^{-i m \gamma t}$, however, since we consider
the momentum $ \vec{p} = m \gamma \vec{v}$ to be fixed, the $\gamma$ in such 
an expression would be $m$ dependent.  
For very narrow resonances, one can consider the difference between $m$ and 
the central value of the resonance mass, $m_0$, to be small and perform 
a Taylor expansion of $\sqrt{p^2 + m^2}$. Noting that the 
fixed momentum $\vec{p}$ is related to the central value of the resonance 
mass, $m_0$, as, $p^2 = E_0^2 - m_0^2 = \gamma^2 m_0^2 - m_0^2$,  
retaining the first two terms of the expansion leads to,
$A_p^{FK}(t) \propto 
\int_{m_{th}}^{\infty} \,dm\, \omega(m) \, e^{-im t/\gamma}$, implying, 
$P_p(t) \approx P_0(t/\gamma)$, 
as in the relativistic time dilation relation. 
Ref. \cite{urbanowski_rel} provides numerical results comparing the 
survival probabilities in (\ref{specialrel}) and (\ref{survival3}) at 
large times using a Breit-Wigner mass distribution, assuming the minimum 
mass of the decay products to be zero.   
Eq. (\ref{survival3}) can also be found in \cite{GiacosaAPPB,
GiraldiHindawi}.  
Providing an analysis with wave packets, the author in \cite{GiacosaAPPB} 
stressed that, ``there is no whatsoever breaking of special relativity, 
but as usual in QM (quantum mechanics), one should specify which kind of 
measurement on which kind of state is performed".

Studying the intermediate time region where the exponential decay 
dominates, Giraldi, in \cite{GiraldiJPhysA, GiraldiEPhysD}, 
approximated the decay laws at rest with superpositions of exponential 
modes via the Prony analysis. The survival
probability $P_p(t)$, was represented
by the transformed form, $P_0(\phi_p(t))$, 
of the survival probability, $P_0(t)$, at rest. The transformed probability 
$P_p(t)$ was determined by choosing the mass distribution, $\omega(m)$, 
to be symmetric with respect to the central value of the resonance mass. 
The author tried to explain the oscillations in the decay laws 
of unstable systems which were observed in an experiment \cite{GSI} but 
not confirmed later \cite{ozturk}. 
 
  
\subsection{Poincar\'e group of special relativistic space-time 
transformations}
Prior to the above derivations, in the 70's and 80's, one finds literature 
where the authors debated about appropriate representations of the Poincar\'e 
group with unstable particles \cite{exner,williams}. For example, in 
\cite{exner}, Pavel Exner starts by assuming that 
the Hilbert spaces ${\cal{H}}_u$ 
and ${\cal{H}}$ of the unstable particle and a larger isolated system 
(consisting of the particle and its decay products) respectively and a 
unitary representation $U$ of the Poincar\'e group ${\cal{H}}$ are given. 
Further, $U_t$ denote the operators which represent the one-parameter 
subgroup of the time translations $U_t = \exp(-iHt)$. The author shows 
that translational invariance of ${\cal{H}}_u$ is not compatible with 
unitarity of the boosts. Proposing a particular choice of ${\cal{H}}_u$, 
the survival amplitude (taking the spin of the unstable particle 
into account) is found to be
\begin{equation}\label{survival5}
A^{P}_p(t) =  \sum_{j = -s}^s \, \int_{m_{th}} ^{\infty} \,dm\,\int_{\R^3} \, 
\frac{d^3p}{2\sqrt{p^2+m^2}} \exp[ -i p\cdot\Lambda(\vec{\beta})x] \, 
|\Psi_j(m,\vec{p})|^2\, , 
\end{equation}
where $m_{th}$ is the threshold mass, 
$\vec{\beta} = \vec{v}/c$ and $\Psi_j(m,\vec{p})$ describes the 
unstable state. With the Lorentz transformation,  
$\Lambda(\vec{\beta})x = (t/\gamma, \vec{0})$, Eq. (\ref{survival5}) takes 
the form, 
\begin{equation}\label{survival6}
A^{P}_p(t) =  \int_{m_{th}} ^{\infty} \,dm\, \omega(m)\, 
\exp{(-i m t /\gamma)} \,  
\end{equation}
with $\omega(m) = \sum_{j = -s}^s \, \int_{\R^3} \,
{d^3p\, (1 / 2\sqrt{p^2+m^2}}) \, |\Psi_j(m,\vec{p})|^2$. Eq. (\ref{survival6}) 
confirms the relativistic time dilation result in (\ref{specialrel}). 
Considering a space-time dependent survival amplitude, 
Alavi and Giunti \cite{alavi} also recovered (\ref{specialrel}). The result
in (\ref{survival6}) seems consistent with the formalisms used to write 
the decay amplitudes of kaon states \cite{Leebook,Srivastava} with the 
replacement of the time $t$ by the proper time $\tau = t/\gamma$ in the 
time evolution operator $\exp(-i H t)$. 
In the same spirit of replacing $t \to \tau$, the amplitude 
(\ref{survival1}), in the 
Fock-Krylov method could 
be written for the decay of relativistic particles as,  
\begin{equation}\label{survival4}
A_p^{FK2}(t) = \int_{\sqrt{p^2+m_{th}^2}}^{\infty} 
\,dE\, \omega(E) \, e^{-iEt/\gamma} \, .
\end{equation}
 
The decay law of moving unstable systems was also studied by Stefanovich 
\cite{stefanovich1996,stefanovich2018} in 
an approach similar to that of Pavel Exner, however, with 
different conclusions. The author studied the decay law of moving particles 
in instant and point forms of Dirac's relativistic dynamics and concluded 
that the quantum corrections depend on the relativistic form of the 
interaction governing the process. The author showed that in a particular 
version of the point form dynamics (PFD), 
the decay law of the moving particle is given by,
$P^{PFD}_p(t) = P_0(\gamma t)$,  
in contradiction with (\ref{specialrel}). 
The author concluded that the point form interaction cannot be responsible for 
particle decays. Constructing an instant form dynamics, however, the author 
obtained exactly the same expression as in (\ref{survival2p}). 

In the sections to follow, we shall study the survival probability as 
given in Eq. (\ref{survival2p}) and compare with the Einstein's time dilation 
formula (\ref{specialrel}) both analytically and numerically for some 
realistic unstable particles. On the way, we shall discover the nuances 
of the relativistic and ultrarelativistic regions of the decay at short 
times, its implications for the time-energy uncertainty relation, the 
deviation of the exponential decay law for particles in motion from that 
of particles at rest and the ``variation" of the standard power law behaviour 
at large times.   

\section{Mathematical properties of the relativistic survival amplitude and probability}\label{math_rel}
In this section we shall study the survival probability as 
given in Eq. \eqref{survival2p}. Related to the density of states, we make the following assumptions:
\begin{enumerate}
\item It has a branch point at $m=\M$, and the asymptotic expansion around this point is such that
\begin{equation}
\om(m)=(m-\M)^\alpha\,q(m),
\end{equation}
where $\alpha>0$ and $q(m)$ is analytic except in isolated points 
which are simple poles, which come in pairs of complex conjugate numbers 
since the density of states is real. {In addition, $q(z)=O(z^{-(1-\al+\rho)})$ when $z\to\infty$, where $\rho>0$ and $z$ is such that $\Re{z}>0$ and $\Im{z}<0$.}
\item The moments of the density of states are well defined, i.e.,
\begin{equation}\label{moments1}
\delta_n\equiv\int_{\M}^\infty dm\,m^n\om(m)<\infty,\quad n=0,1,2,3,\dotsc
\end{equation}
and for $n=0$, $\delta_0=1$, which is just the normalization condition.
Eq. (\ref{moments1}) is in principle an assumption 
which is necessary to recover the correct expectation values 
\cite{ourpaperPRA}. Thus, $\delta_n$ can be considered to be 
the expectation value of $m$ raised to the power
$n$.  
\item The calculations that we shall perform are under the assumption of the 
{\it dominant pole approximation}, i.e., we take into account the pole of 
$\om(m)$ on the fourth quadrant such that it has the smallest imaginary part only, 
and we neglect the contributions of the remaining poles. 
Hence, let the complex number $m_0-i\,\dfrac{\Ga_0}{2}$ be the dominant pole, 
where $m_0>\M$ and $\Ga_0>0$.
\end{enumerate}
In the nonrelativistic treatment of the decay of unstable systems, it is customary to write the survival amplitude in terms of the parameter $\dfrac{\Ga_0}{2m_0}$. However, in that context, $m_0$ is supposed to be $m_0-\M$ and not $m_0$ because a change of variable $m\to m-\M$ is made such that the lower limit of integration will be zero, and as a consequence, $m_0$ is implicity assumed to be $m_0-\M$. Henceforth, not only would we like to introduce the same parameter, that is,
\begin{equation}
x_0=\frac{\Ga_0}{2(m_0-\M)},
\end{equation}
but also to redefine the time as the dimensionless quantity
\begin{equation}
\tau=\Ga_0 t,
\end{equation}
which are defined in the same way as in the nonrelativistic formalism. Thus, the survival amplitude given by the eq. \eqref{survival2p} transforms as:
\begin{equation}
A_p^{FK}(\tau)=\int_{\M}^{\infty}{dm\,\om(m)\,\exp{\GL{-\frac{i\tau}{2x_0}\,\frac{\sqrt{p^2+m^2}}{m_0-\M}}}}.
\end{equation}
If we make the change of variable $m=(m_0-\M)\xi$, that is,
\begin{equation}
A_p^{FK}(\tau)=(m_0-\M)
\int_{\frac{\M}{m_0-\M}}^{\infty}{
d\xi\,
\om((m_0-\M)\xi)
\exp{
\EL{
-\frac{i\tau}{2x_0}\,\sqrt{\xi^2+\GP{\frac{p}{m_0-\M}}^2}
}
}
},
\end{equation}
and if we introduce the additional parameters
\begin{align}\label{Ppara}
\mu&=\frac{\M}{m_0-\M},\\
P&=\frac{p}{m_0-\M},
\end{align}
which let us rewrite the factor $m-m_0+i\dfrac{\Ga_0}{2}$ responsible for the pole as
\begin{equation}
m-m_0+i\dfrac{\Ga_0}{2}=(m_0-\M)(\xi-1-\mu+ix_0),
\end{equation}
allows us to write the density of states as
\begin{equation}
\lam(\xi)\equiv (m_0-\M)\om((m_0-\M)\xi).
\end{equation}
This density of states will have a branch point in $\xi=\mu$, its dominant pole will be $\xi=1+\mu-ix_0$, and since $\om(m)=(m-\M)^\al\,q(m)$, $\lam(\xi)$ will be decomposed as
\begin{equation}
\lambda(\xi)=(m_0-\M)\cdot(m-\M)^\alpha\,q(m)=(\xi-\mu)^\al\cdot\underbrace{(m_0-\M)^{\al+1}q\NP{(m_0-\M)\xi}}_{\equiv Q(\xi)}=(\xi-\mu)^\al Q(\xi),
\end{equation}
and thus $Q(\xi)$ will have a pole in $\xi=1+\mu-ix_0$ and  $z\to\infty$ on 
the fourth quadrant, $Q(z)=O(z^{-(1-\al+\rho)})$. As a result, the survival amplitude reads:
\begin{equation}\label{eq-math7}
a(\tau,P)\equiv A_p^{FK}(\tau)=
\int_{\mu}^{\infty}
{d\xi\,\lam(\xi)\,\exp{\EP{-\frac{i\tau}{2x_0}\,\sqrt{\xi^2+P^2}}}}.
\end{equation}
For future references, the moments of the density of states are rewritten as:
\begin{equation}\label{momenta_H}
\tilde{\delta}_n=\GP{\frac{1}{2x_0}}^n
\int_\mu^\infty\,d\xi\,\xi^n\lam(\xi) \, =\, \frac{1}{\Gamma_0^n} \, 
\delta_n<\infty,\quad n=0,1,2,\dotsc.
\end{equation}
Because of the comparison test for improper integrals, the integral \eqref{eq-math7} converges for $\tau\geq0$ and $P\geq0$ provided that the integral of the density of states over the interval $\xi\geq\mu$ also converges, namely,
\[
\EB{\int_{\mu}^{\infty}{d\xi\,\lam(\xi)\,\exp{\EP{-\frac{i\tau}{2x_0}\,\sqrt{\xi^2+P^2}}}}}\leq\int_{\mu}^{\infty}{d\xi\,\NB{\lam(\xi)}}=\int_{\mu}^{\infty}{d\xi\,{\lam(\xi)}}.
\]
Recall that the density of states is positive and its integral over the interval $\xi\geq\mu$ exists thanks to the condition \eqref{momenta_H}. Moreover, because of the Weierstrass test \cite{Titch}, the integral \eqref{eq-math7} converges uniformly in the domain $0\leq \tau\leq \tau_0$ and $0\leq P\leq P_0$, where $\tau_0$ and $P_0$ are arbitrary. This feature implies that we can only take the limit when $P$ tends to infinity once the integral is computed.

Finally, we also add that the convergence of (\ref{momenta_H}) 
is not necessarily guaranteed by that of $\lambda(\xi)$,
but by the form factor hidden inside the particular form of $\lambda(\xi)$. 
The latter is mostly chosen to be of an exponential form which is strong 
enough to ensure convergence of (\ref{momenta_H}). 
\subsection{Non--relativistic survival amplitude}\label{NRsurv}
Since the nonrelativistic survival amplitude is rather important throughout the future discussions, this section will summarize the main results regarding 
the topic. The proofs of all of those results are shown in 
\cite{ourpaperPRA}.

The non--relativistic survival amplitude is obtained when we put $P=0$ in Eq. \eqref{eq-math7}:
\begin{equation}\label{eq-math8}
a(\tau,P=0)=\int_{\mu}^{\infty}
{d\xi\,\lam(\xi)\,\exp{\EP{-\frac{i\tau}{2x_0}\,\xi}}}.
\end{equation}
It is well known that the nonrelativistic survival amplitude can be decomposed as a sum of an exponential and a non-exponential component:
\begin{equation}\label{eq-math9}
a(\tau,P=0)=a_e(\tau,P=0)+a_{ne}(\tau,P=0),
\end{equation}
where these components are given in the dominant pole approximation as
\begin{align}
a_e(\tau,P=0)&=-2\pi i\Res{\NC{\lam(z),z=\mu+\zeta_0}}e^{-i\mu\tau/2x_0}
\exp{\EC{-\GP{\frac{1}{2}+i\frac{1}{2x_0}}\,\tau}},\label{eq-math10}\\
a_{ne}(\tau,P=0)&=\int_{\mu}^{\mu-i\infty}{dz\,\lam(z)\,\exp{\EP{-i\frac{\tau}{2x_0}\,z}}}
\notag\\&=-ie^{-i\mu\tau/2x_0}
\int_{0}^{\infty}{d\eta\,\lam(\mu-i\eta)\,\exp{\EP{-\frac{\tau}{2x_0}\,\eta}}}
,\label{eq-math11}
\end{align}
where $\zeta_0=1-ix_0$. For large times, that is, $\dfrac{\tau}{2x_0}\gg1$, the non-exponential survival amplitude approximates as
\begin{equation}\label{eq-math13}
a_{ne}(\tau,P=0)\sim\Ga\NP{\al+1}\,e^{-i\mu\tau/2x_0}\,Q(\mu)\GP{\frac{2x_0}{i\tau}}^{\al+1}.
\end{equation}
In addition, the survival amplitude for small times follows a quadratic law in $\tau$, that is,
\begin{equation}\label{stimes_nr}
S(\tau,P=0)=1-(\tilde{\delta}_2-\tilde{\delta}_1^2)\tau^2+O\NP{\tau^4}
=1-({\delta}_2-{\delta}_1^2)t^2+O\NP{t^4},
\end{equation}
where, $(\tilde{\delta}_2-\tilde{\delta}_1^2)\,\Gamma_0^2={\delta}_2-{\delta}_1^2$, 
is the square of the uncertainty of the nonrelativistic Hamiltonian at the initial state of the unstable system.
In passing, we note that when the spectral function becomes very narrow (a 
Dirac delta in the extreme case),
$(\tilde{\delta}_2-\tilde{\delta}_1^2)$, $(\tilde{\delta}_4-
\tilde{\delta}_2^2)$ and similar expressions go to zero.  
Now, we define the nonrelativistic survival probability $S(\tau,P=0)$ as the modulus square of the survival amplitude, that is:
\begin{equation}
S(\tau,P=0)\equiv|a(\tau,0)|^2=|a_e(\tau,0)|^2+|a_{ne}(\tau,0)|^2+2\Re{\MC{a_e(\tau,0)a_{ne}^*(\tau,0)}}.
\end{equation}
$S(\tau,P=0)$ has the following properties:
\begin{enumerate}[i)]
\item $S(\tau,P=0)$ is oscillatory, and the respective angular frequency is 
determined by the frequency of oscillation of the function $\dfrac{a_e(\tau,0)}{a_{ne}(\tau,0)}$. Such a frequency is just $\dfrac{1}{2x_0}$.

\item The critical time for large times $\tau_{lt}$ is defined as the largest solution of the equation
\begin{equation}
|a_e(\tau,0)|^2=|a_{ne}(\tau,0)|^2.
\end{equation}
One remarkable feature about this critical time is that the smaller $x_0$ is, the larger the critical times is.
\item The critical time for small times $\tau_{st}$ is defined when the survival probability has completed one oscillation \cite{ourpaperPRA}, that is,
\begin{equation}
\tau_{st}=\frac{2\pi}{1/2x_0}=4\pi x_0.
\end{equation}
This time indicates when the survival probability starts to be dominantly exponential, and smaller the value of $x_0$ is, the principal contribution to the survival probability at that time will come from its exponential component.
\end{enumerate}
\subsection{Exponential decay at intermediate times}\label{intert}
We shall now calculate the exponential component of the relativistic survival amplitude. For now, let us define this component as $-2\pi i$ times the residue of the density of states at its pole on the fourth quadrant times the relativistic phase factor without specifying any integration contour yet:
\begin{align}
a_{e}(\tau,P)
&=-2\pi i\Res{\EC{{\lam(z)}\,\exp{\EP{-\frac{i\tau}{2x_0}\,\sqrt{z^2+P^2}}},z=1+\mu-ix_0}}\notag\\
&=R\exp{\EC{-\frac{i\tau}{2x_0}\,{\sqrt{(\zeta_0+\mu)^2+P^2}}}},
\end{align}
where $\zeta_0=1-ix_0$ and $R=-2\pi i\Res{\NC{\lam(z)\,,z=\mu+\zeta_0}}$. If we denote the real and imaginary part of $\sqrt{(\zeta_0+\mu)^2+P^2}$ respectively as $2 x_0 \Omega$ and $-x_0 \sigma$,
\begin{equation}
a_{e}(\tau,P)=R\,\exp{\EC{-\GP{\frac{\sigma}{2}+i{\Omega}}\,\tau}},
\end{equation}
and we identify $\sigma$ as the decay rate of the exponential component of the survival probability and $\Omega$ as the frequency of oscillation of the exponential survival amplitude. In the Appendix \ref{app2}, we prove that $\sigma$ is positive when $x_0>0$. Moreover, in the same appendix we deduce that
\begin{align}
\Omega&=
\frac{1}{2\sqrt{2}x_0}\GC{\sqrt{\NC{(P+x_0)^2+(1+\mu)^2}\NC{(P-x_0)^2+(1+\mu)^2}}+P^2+(1+\mu)^2-x_0^2}^{1/2},
\\
\sigma&=
\frac{1}{\sqrt{2}x_0}\GC{\sqrt{\NC{(P+x_0)^2+(1+\mu)^2}\NC{(P-x_0)^2+(1+\mu)^2}}-P^2-(1+\mu)^2+x_0^2}^{1/2}.
\end{align}
Since $\sigma\cdot\Omega=\dfrac{1+\mu}{2x_0}$, they are not independent quantities as in the nonrelativistic case. What is more, the larger (smaller) the decay rate, the smaller (larger) the frequency of oscillation.

For $P=0$, $\Omega$ and $\sigma$ reduce to:
\begin{align}
\Omega\Bigr|_{P=0}&=\frac{1+\mu}{2x_0},\\
\sigma\Bigr|_{P=0}&=1.
\end{align}
Both expressions reduce to the nonrelativistic case as 
Eq. \eqref{eq-math10} shows.

One of the advantages of introducing the parameters $x_0$, $\mu$ and  $P$ is to study how $\sigma$ and $\Omega$ behave when the momentum of the system is large, and what a large momentum in an unstable, relativistic system means. If $x_0$ is negligible with respect to $P$, but $1+\mu$ is not, $\Omega$ and $\sigma$ take the following forms:
\begin{align}
\Omega\Bigr|_{P\gg x_0}&\approx\frac{\sqrt{P^2+(1+\mu)^2}}{2x_0},\\
\sigma\Bigr|_{P\gg x_0}&\approx\frac{1+\mu}{\sqrt{P^2+(1+\mu)^2}}.
\end{align}
If, in the former case, $1+\mu$ is negligible with respect to $P$ as well:
\begin{align}
\Omega\Bigr|_{\substack{P\gg x_0\\ P\gg 1+\mu}}&\approx\frac{P}{2x_0},\\
\sigma\Bigr|_{\substack{P\gg x_0\\ P\gg 1+\mu}}&\approx \frac{1+\mu}{P}.
\end{align}
Summarizing, we can say that $P$ is large if $P\gg 1+\mu$, and $P\gg x_0$. Moreover, $\Omega=O(P)$ and $\sigma=O(P^{-1})$ for large $P$. Physically, this means that large momenta imply that the system decays slowly and therefore the 
decay will be dominantly exponential. Another implication of this result is 
that the time taken by the relativistic survival probability to both leave 
the quadratic, small time regime and enter from the exponential into 
the non--exponential, large time regime will be larger than its 
nonrelativistic counterpart {\it provided that the momentum is large}. 
We shall return to this point in section \ref{c_times}. Finally, 
we show in Table \ref{tab21} the values of the parameters 
$x_0$, $\mu$ and $P$ for some decay processes.

\begin{table}[htb!]
\centering
\begin{tabular}{c|c|c||c|c|c|c}\cline{4-7}
\multicolumn{3}{c}{} & $p=0.5$ & $p=1$ & $p=5$ & $p=10$ \\\hline
Process & $x_0$ & $\mu$ & $P$ &  $P$ & $P$ & $P$ \\\hline
$\Delta^{++}\to p+\pi^+$ & 0.37987  & 7.0000 & 3.2468 & 6.4935 & 12.987 & 64.935\\\hline
$\rho^0\to\pi^++\pi^-$ & 0.15100  & 0.56131 & 1.0054 & 2.0107 & 4.0214 & 20.107 \\\hline
$Z^0$ & $1.3684\times10^{-5}$ & $1.1208\times10^{-5}$ & 0.0054833 & 0.010967 & 0.02193 & 0.10967\\\hline
$\mu^-\to e^{-}+\bar{\nu}_e+\nu_\mu$ & $1.4218\times10^{-18}$ & $4.8598\times10^{-3}$ & 4.7552 & 9.5103 & 19.021 & 95.103 \\\hline
$K^+\to\mu^++\nu_\mu$ & $6.8554\times10^{-17}$ & 0.27231 & 1.2886 & 2.5772 & 5.1544 & 25.772\\\hline
\end{tabular}
\caption{Values of the parameters for some examples of broad and narrow decays
as well as the parameter, $P=p/(m_0-m_\text{th})$, defined in Eq. (\ref{Ppara}) for these decays at different momenta, $p$, given in GeV. Data for each process is available in \cite{pdg}.}\label{tab21} 
\end{table}

Moreover, let us consider the relativistic, exponential, survival probability $P_e(\tau,P)$:
\begin{equation}
P_e(\tau,P)=|a_e(\tau,P)|^2=|R|^2\,e^{-\sigma\,\tau},
\end{equation}
and their counter--nonrelativistic part:
\begin{equation}\label{nonrelexp}
P_e(\tau,0)=|R|^2\,e^{-\tau},
\end{equation}
We note that in general, $|R|^2 \ne 1$ when we normalize over the whole time 
region including small and large times. After the first measurement in the 
exponential region, one can take $|R|^2 = 1$. 
From the Eq. \eqref{nonrelexp} and with the help of the Eq. \eqref{specialrel} we obtain the survival probability of the unstable particle in motion $p_e(\tau,\ga)$, that is,
\begin{equation}
p_e(\tau,\ga)=|R|^2\,\exp{\GP{-\frac{\tau}{\gamma}}}.
\end{equation} 
This is essentially the result we expect from time dilation arguments. 
We would like to compare the ratio $\dfrac{P_e(\tau,P)}{p_e(\tau,\ga)}$ as a 
function of $\gamma$ in order to see how much $P_e(\tau,P)$ deviates from $p_e(\tau,\ga)$. From the definitions of the respective survival probabilities:
\begin{equation}\label{ratioexponentials}
\frac{P_e(\tau,P)}{p_e(\tau,\ga)}=\exp{\MC{-\MP{\sigma(P,x_0)-\gamma^{-1}}\tau}}\,.
\end{equation}
Note that in the ultrarelativistic (UR) 
limit, $p = m_0 \gamma v = 
m_0 \sqrt{\gamma^2 -1 } \simeq m_0 \gamma$ and using the definition of 
$\sigma \approx (1 + \mu)/P$, 
\begin{equation}
P^{UR}_e(\tau,P) = |R|^2\,\exp{\GP{-\frac{\tau}{\gamma}}}
\end{equation}
and the ratio in (\ref{ratioexponentials}) is unity. This can be seen 
as follows too:
since the exponential function is a decreasing, monotonous function, and in addition $\tau>0$, the properties of this ratio follow from the properties of the function
\begin{equation}
f(\gamma,x_0)=\sigma(P,x_0)-\ga^{-1}
\end{equation}
in terms of $\gamma$, which is done in the Appendix \ref{app3}. The analysis of this function reveals that:
\begin{enumerate}[i)]
\item {\it For constant $x_0$}, it turns out that $f(\infty,x_0)=0$, and as a consequence,
\begin{equation}
\lim_{\ga\to\infty}\frac{P_e(\tau,P)}{p_e(\tau,\ga)}=1.
\end{equation}
Moreover, under the same conditions, $f(\gamma,x_0)$ has a maximum in $\gamma=\gamma_c$ given by
\begin{multline}\label{gammac}
{\ga_c^2}=\frac{5}{3}+\frac{9}{25}\GP{\frac{x_0}{1+\mu}}^2
-\frac{31}{3125}\GP{\frac{x_0}{1+\mu}}^4\\
+\frac{267}{390625}\GP{\frac{x_0}{1+\mu}}^6-\frac{573}{9765625}\GP{\frac{x_0}{1+\mu}}^8+\frac{33642}{6103515625}\GP{\frac{x_0}{1+\mu}}^{10}+\cdots.
\end{multline} 
Hence, we conclude that the ratio $\dfrac{P_e(\tau,P)}{p_e(\tau,\ga)}$ has a minimum when {$\ga_c^2$} is given by this power series, that is, for this value of $\gamma$ we have the maximum deviation between $P_e(\tau,P)$ and $p_e(\tau,\ga)$. Notice that, for narrow resonances, this minimum is such that ${\ga_c^2}
\approx\dfrac{5}{3}$. Numerical tests, shown in Fig. \ref{ratio}, 
validate this result, and it demonstrates that similar calculations (see \cite{khalfinRel}) where this maximum is such that $\ga_c^2=1+4\GP{\dfrac{x_0}{1+\mu}}^2$ is incorrect.

\begin{figure}[htb!]
\centering
\includegraphics[scale=0.45]{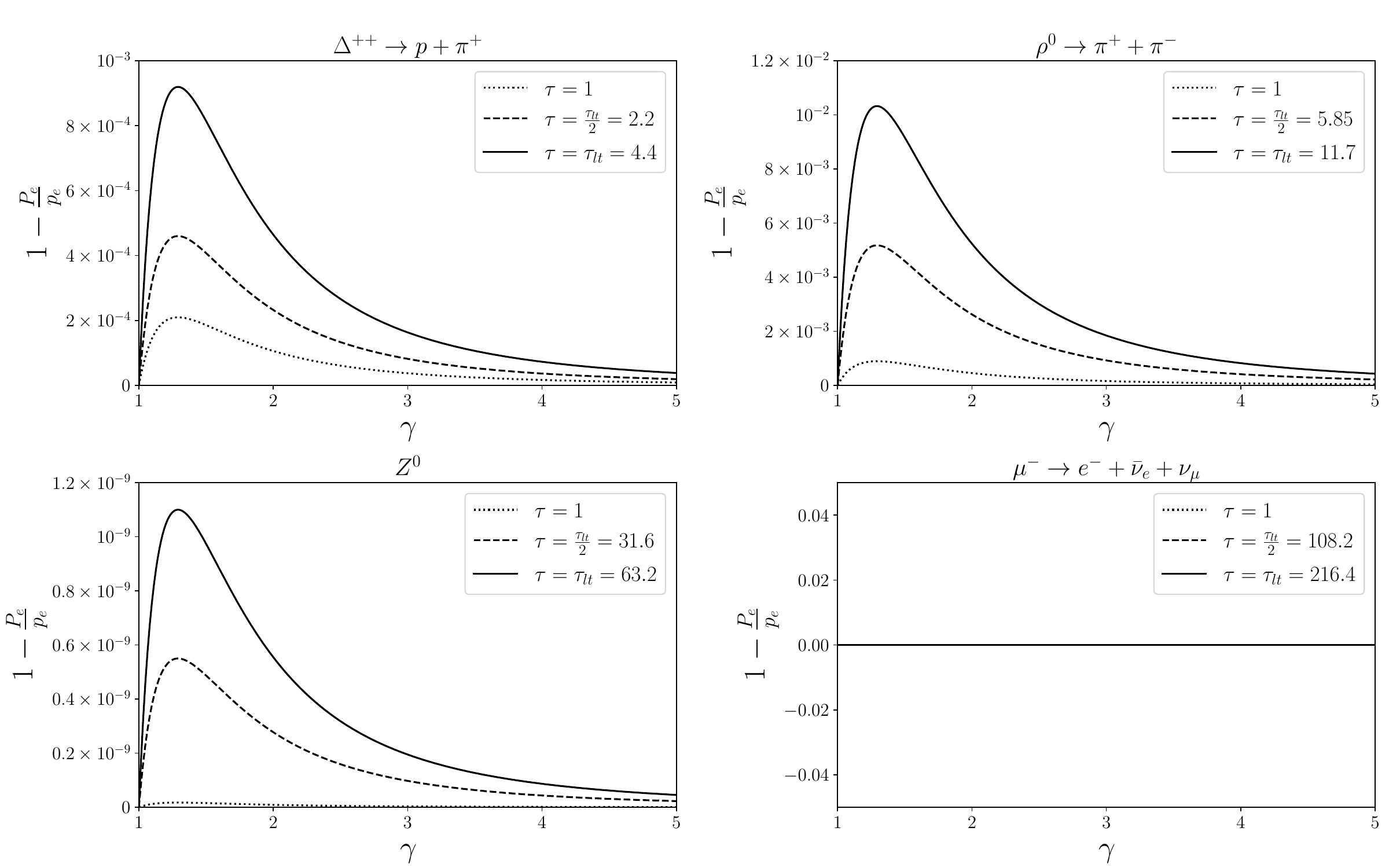}
\caption{Ratio of the relativistic survival probability, $P_e(\tau,P)$ and 
$p_e(\tau,\ga)$, obtained using the time dilation formula, for different 
broad and narrow resonances as a function of the Lorentz factor, $\gamma$. The values for $P$ and $x_0$ used for each resonance are shown in table \ref{tab21}.
Different curves in each panel are drawn at times corresponding to 
approximately the beginning of the exponential region ($\tau =1$), half of the critical 
transition time, $\tau_{lt}/2$ and at $\tau_{lt}$. The values of $\tau_{lt}$ 
are chosen to be those of the decays at rest though the behaviour of
the curves may not change significantly for the corresponding values in 
flight.}  
\label{ratio}
\end{figure}

\item {\it For constant $\gamma$}, $f(\gamma,0)=0$, and as a result,
\begin{equation}
\lim_{x_0\to0}\frac{P_e(\tau,P)}{p_e(\tau,\ga)}=1.
\end{equation}
In addition, $\sigma(\gamma,x_0)$ (and so does $f(\ga)$) behaves as a 
monotonic increasing function for $x_0$ if $0<x_0<1$. The consequence of this property is that the ratio $\dfrac{P_e(\tau,P)}{p_e(\tau,\ga)}$ decreases monotonically when $x_0$ goes from 0 to 1 for a constant $P$ (or constant $\ga$), or 
paraphrasing this statement, ${P_e(\tau,P)}$ deviates from ${p_e(\tau,\ga)}$ the most for broad resonances.
\end{enumerate}
\subsection{Non-exponential decay at large times}\label{larget}
\begin{figure}[htb!]
\centering
\includegraphics[scale=1.0]{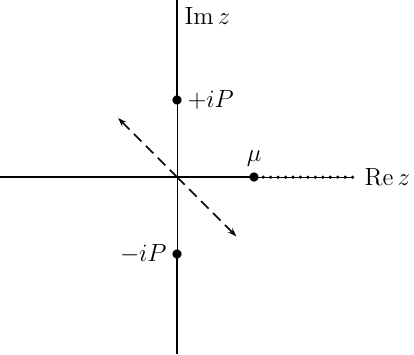}
\caption{Complex plane where the integral \eqref{eq-math7} is studied.}\label{f1}
\end{figure}

In this section we shall study the asymptotic behavior of the relativistic survival probability given by Eq. \eqref{eq-math7} when $\dfrac{\tau}{2x_0}\gg1$. Let us write this integral as a complex contour integral, that is,
\begin{equation}\label{eq-lt1}
a(\tau,P)=
\int_{\mu}^{\infty}
{
dz\,
\lam(z)\,
\exp{
\MP{
-ik\,\sqrt{z^2+P^2}
}
}
},
\end{equation}
where $k$ is $\dfrac{\tau}{2x_0}$, and the contour of integration starts in the point $z=\mu$ and goes along the positive real axis. The contour is represented in figure \ref{f1} as a dotted line. Here, the complex plane (actually, cut complex plane) is such that it has three branch points in $\pm iP$ and $\mu$, and the cuts are chosen as we show in the figure \ref{f1}.

In addition, we are interested in the saddle points (we follow the notation and language from \cite{Ablowitz}) of the integral \eqref{eq-lt1}. Since we want to calculate this integral for large $k$, the saddle points will be given by the function $\phi(z)=-i\sqrt{z^2+P^2}$. Thus, let us calculate its first and second derivatives:
\begin{align}
\phi'(z)&=-\frac{iz}{\sqrt{z^2+P^2}},\label{eq-lt2}\\
\phi''(z)&=-\frac{2iP^2}{(z^2+P^2)^{3/2}}.\label{eq-lt3}
\end{align}
These equations reveal that $\phi'(z)=0$ when $z=0$, but the second derivative is nonzero, and it has as a value $\phi''(0)=\dfrac{1}{P}e^{-i\pi/2}$. Hence, the integrand has a simple saddle point at $z=0$, and the steepest descent 
directions are $\dfrac{3}{4}\pi$ and $-\dfrac{1}{4}\pi$ which are 
represented with dashed lines in figure \ref{f1}.

Since the saddle point is at $z=0$, the contour might be deformed so that it would go through the saddle point and it would be tangent at that point. However we would be forced to include another integral along the segment over the real axis from zero to $\mu$, and since the lower limit of integration is the saddle point, we would have to compute two contributions for large $k$.

\begin{figure}[htb!]
\centering
\includegraphics[scale=1.5]{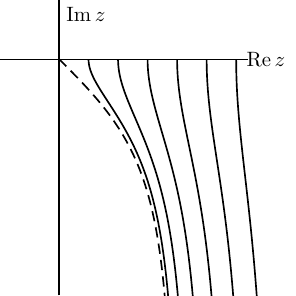}
\caption{Descent paths for the function $-i\sqrt{z^2+P^2}$. The dashed line corresponds to $\mu=0$ (steepest descent), and the dark, solid lines correspond to 
$\mu\neq0$.}\label{f2}
\end{figure}

Instead of forcing the current contour of integration to be deformed along the saddle point, it is better to seek a contour of integration in which the imaginary part of $\phi(z)$ is constant, and therefore the integrand will not contain highly oscillatory terms which depend on $z$ (recall that for large $k$, the function $e^{-ik\phi(z)}$ is highly oscillatory). In other words, we shall find 
the paths of descent of the integral \eqref{eq-lt1}.
Moreover, the point $z=\mu$ belongs to this contour, so that the equation that describes the paths of descent is given by
\begin{equation}\label{eq-lt5}
\Im{\NP{-i{\sqrt{z^2+P^2}}}}=-{\sqrt{\mu^2+P^2}}.
\end{equation}
Let $u$ and $v$ be the real and imaginary part of $-i{\sqrt{z^2+P^2}}$, with $v=-{\sqrt{\mu^2+P^2}}$ and if $z=\xi+i\eta$,
\[
(u+iv)^2=u^2-v^2+2iuv=-(\xi^2-\eta^2+P^2+2i\xi\eta).
\] 
Comparing the real and imaginary parts of the above equation:
\begin{align}
u^2-v^2&=-(\xi^2-\eta^2+P^2),\label{eq-lt6}\\
uv&=-\xi\eta.\label{eq-lt7}
\end{align}
Since we are interested in the paths on the fourth quadrant, $\xi>0$, $\eta<0$ and $v<0$, and thanks to Eq. \eqref{eq-lt7}, $u<0$. Eliminating $u$ from Eqs \eqref{eq-lt6} and \eqref{eq-lt7}, and taking into account that $v^2=\mu^2+P^2$, we find the cartesian equations for the descent paths:
\begin{equation}\label{eq-lt8}
\GP{1+\frac{\eta^2}{v^2}}\GP{1-\frac{\xi^2}{v^2}}=\frac{P^2}{v^2}.
\end{equation}
Notice that these paths are symmetrical with respect to the axes $\xi$ and $\eta$, and also the origin. These loci have two properties, which are easy to demonstrate. The first property tells us that the loci have asymptotes at $\Re{z}=\sqrt{\mu^2+P^2}$, and the second property is that the angle between each locus and the real axis, i.e., $\Re{z}=\mu$, is $\pi/2$. However, the angle between the locus associated with $\mu=0$ (called as the bullet nose, see \cite{Lawrence}), whose cartesian equation is given by 
\begin{equation}\label{eq-lt9}
\frac{1}{\xi^2}-\frac{1}{\eta^2}=\frac{1}{P^2},
\end{equation}
and the real axis is $\pi/4$. This is to say, the direction of the tangent in 
the origin is no longer perpendicular to the real axis, 
but is along the bisectrix of the fourth quadrant (see figure \ref{f2}), 
which is nothing but one of the 
steepest descent directions.

\begin{figure}[htb!]
\centering
\includegraphics{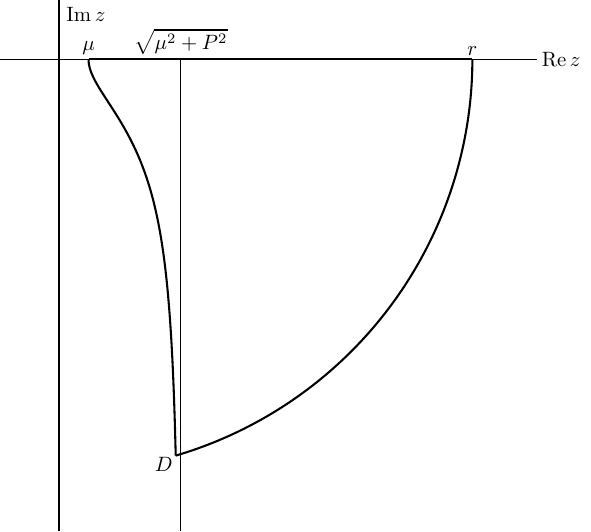}
\caption{Deformation of the contour of integration of \eqref{eq-lt1} along the descent path.}\label{f3}
\end{figure}

The interesting feature about the angles between the descent paths and the real axis at their intersection points with the real axis is that they change discontinuously when $\mu$ is zero. This has a consequence in the asymptotic expansion of the integral \eqref{eq-lt1} for large $k$, that is, the asymptotic expansion will show a discontinuity for $\mu=0$ but we cannot realize the nature of this discontinuity once the asymptotic expansion will be calculated. Regardless whether $\mu$ is zero or not, we know that in these contours, the imaginary part of $\phi(z)$ is constant, and since its real part is negative over these paths, it opens the gate to apply the Watson's lemma \cite{Ablowitz}. 
In order to do so, let us modify the contour of integration as is shown in Fig. \ref{f3}. 

The contour of integration consists of: i) a segment over the positive real 
axis starting from $z=\mu$ to $z=r>\sqrt{\mu^2+P^2}$, ii) an arc of 
circumference $C_r$ of radius $r$ centered at $z=0$, and this arc goes 
clockwise from $z=r$ until the arc intersects the descent path in the point 
$D$, and iii) the segment of the descent path $\Gamma$ from $D$ to $z=\mu$.

Assuming that the pole $z=1+\mu-ix_0$ is inside the contour, which happens if
\begin{equation}
\NP{P^2+\mu^2}\,\frac{2\mu+1}{P^2-\NP{2\mu+1}}>x_0^2,
\end{equation}
from the residue theorem we get that
\begin{equation}\label{eq-lt10}
\int_{\mu}^{r}+\int_{C_r}+\int_\Gamma=R\,e^{-\sigma\,\tau/2}\,e^{-i\tau\,\Omega},
\end{equation} 
where the dominant pole approximation \cite{ourpaperPRA} was used once again. 
Assuming that the Jordan's Lemma is satisfied over $C_r$, and making $r\to\infty$, we obtain:
\begin{equation}\label{eq-lt11}
a(\tau,P)=R\,e^{-\sigma\,\tau/2}\,e^{-i\tau\,\Omega}+
\int_{\mu}^{\sqrt{\mu2+P^2}-i\infty}{dz\,\lam(z)\,\exp{\NP{-ik\,\sqrt{z^2+P^2}}}},
\end{equation}
where the last integral goes along the descent path. The descent path is parametrized through the change of variable $-i\sqrt{z^2+P^2}=-s-i\sqrt{\mu^2+P^2}$, with $s>0$ (recall that the real part of $-i\sqrt{z^2+P^2}$ is negative along the descent path).
As a result, the integral in Eq. \eqref{eq-lt11} is
\begin{multline}\label{eq-lt12}
\int_{\mu}^{\sqrt{\mu2+P^2}-i\infty}{dz\,\lam(z)\exp{\NP{-ik\,\sqrt{z^2+P^2}}}}\\
=-e^{-ik\sqrt{\mu^2+P^2}}\int_{0}^{\infty}ds\,e^{-ks}\,\frac{s+i\sqrt{\mu^2+P^2}}{\sqrt{\mu^2-2is\sqrt{\mu^2+P^2}-s^2}}\,\lam\MP{\sqrt{\mu^2-2is\sqrt{\mu^2+P^2}-s^2}},
\end{multline}
where the real and imaginary parts of $z=\sqrt{\mu^2-2is\sqrt{\mu^2+P^2}-s^2}$ are
\begin{multline}
\sqrt{\mu^2-2is\sqrt{\mu^2+P^2}-s^2}
=\sqrt{\frac{\sqrt{(\mu^2+s^2)^2+4P^2s^2}+\mu^2-s^2}{2}}\\
-i\,\sqrt{\frac{\sqrt{(\mu^2+s^2)^2+4P^2s^2}-\mu^2+s^2}{2}},\quad s>0.
\end{multline}
Finally, in order to obtain the asymptotic expression for this integral for large $k$, all we need to do is an expansion around $s=0$ in order to find the leading contribution. A first inspection of the integral reveals that the discontinuity we mentioned around $\mu=0$ is hidden in the term $\sqrt{\mu^2-2is\sqrt{\mu^2+P^2}-s^2}$  because for $\mu\neq0$ the leading term around $s=0$ is 
just $\mu$, and for $\mu=0$ the leading terms will be proportional to 
$s^{1/2}$. Accordingly,
\begin{equation}\label{eq-lt12a}
\sqrt{\mu^2-2is\sqrt{\mu^2+P^2}-s^2}=
\begin{cases}
\mu-i\sqrt{1+\dfrac{P^2}{\mu^2}}\,s+\dfrac{P^2}{2\mu^3}\,s^2+\cdots & 
\mu\neq0 \,, \\
(1-i)(Ps)^{1/2}-\dfrac{1+i}{4\sqrt{P}}s^{\,3/2}+\cdots  & \mu=0 \,. 
\end{cases}
\end{equation}
This implies that the discontinuity will be contained in the exponent of the leading term in the asymptotic expansion of the integral, so that we need to compute two expansions around $s=0$: one of them is for $\mu\neq0$ and another one for $\mu=0$. Recalling that
\begin{equation}
\lam(\xi)=(\xi-\mu)^\alpha Q(\xi),
\end{equation}
we have for the former case that the integrand (let us call it $I$) is given by
\begin{equation}\label{eq-lt13}
I=-e^{-ks}\,e^{-ik\sqrt{P^2+\mu^2}}\,\EL{Q(\mu)\EP{-i\sqrt{1+\frac{P^2}{\mu^2}}\,}^{\al+1}s^{\al}+O(s^{\al+1})};
\end{equation}
and for the latter case it is given by
\begin{equation}\label{eq-lt14}
 I=-e^{-ks}\,e^{-ikP}\,\EL{Q(0)(-2iP)^{(\al+1)/2}s^{(\al-1)/2}+O(s^{(\al+1)/2})}.
\end{equation} 
Calling the integral along the descent paths $A_{lt}(k)$, 
($lt\equiv\text{large time}$) the asymptotic expansion for large $k$ follows from the Watson's lemma\footnote {The Watson's lemma allows to compute integrals of the form $\int_0^b{f(t)e^{-kt}\,dt}$ when $k$ is large. Suppose that $f(t)$ has the asymptotic series expansion
\[
f(t)\sim t^\al\sum_{n=0}^{\infty}{a_nt^{\,\beta n}},\quad t\to0^+,\quad\al>-1,\beta>0.
\]
Then
\[
\int_0^b{f(t)e^{-kt}\,dt}\sim\sum_{n=0}^{\infty}{a_n\,\frac{\Gamma\NP{\al+\beta n+1}}{k^{\,\al+\beta n+1}}},\quad k\to\infty.
\]
For $t>0$: If $b$ is finite, it is required that $|f(t)|<A$, where $A$ is a constant; if $b=\infty$, $f(t)$ must be a function of exponential order, that is, $|f(t)|<Me^{ct}$ with $c$ and $M$ are constants \cite{Ablowitz}.}:
\begin{equation}\label{NElargeRel1}
A_{lt}(\tau)\sim
\begin{cases}
-Q(\mu)\EP{-i\sqrt{1+\dfrac{P^2}{\mu^2}}\,}^{\al+1}\,e^{-ik\sqrt{P^2+\mu^2}}\,\dfrac{\Gamma(\al+1)}{k^{\al+1}} & \mu\neq0 \,,\\
 & \\
-Q(0)(-2iP)^{(\al+1)/2}\,e^{-ikP}\,\dfrac{\Gamma\NP{\frac{\al+1}{2}}}{k^{(\al+1)/2}} & \mu=0 \, ,
\end{cases}
\end{equation}
where $\sim$ means that the expression is {\it asymptotically} equal to. In terms of $\tau$, the survival amplitude for large times reads:
\begin{equation}\label{NElargeRel2}
A_{lt}(\tau)\sim
\begin{cases}
-Q(\mu)\EP{-2ix_0\sqrt{1+\dfrac{P^2}{\mu^2}}\,}^{\al+1}\,e^{-i\tau\sqrt{P^2+\mu^2}/2x_0}\,\dfrac{\Gamma(\al+1)}{\tau^{\al+1}} & \mu\neq0 \,,\\
 & \\
-Q(0)(-4ix_0P)^{(\al+1)/2}\,e^{-i\tau P/2x_0}\,\dfrac{\Gamma\NP{\frac{\al+1}{2}}}{\tau^{(\al+1)/2}} & \mu=0 \, .
\end{cases}
\end{equation} 
In all of the calculations for large times we have not considered the nonrelativistic case ($P=0$). For $\mu\neq 0$, if we consider $P\to 0$, 
the descent paths go over to the line $\Re{z}=\mu$ continuously, which can be obtained from the Eq. \eqref{eq-lt8}, and we should be able to recover the nonrelativistic case from the first formula in Eq. \eqref{NElargeRel2}. For $\mu=0$, we already know that the angle between the descent paths and the real axis at the origin of the z--complex plane is always $\pi/4$ for any nonzero momentum. However, Eq. \eqref{eq-lt8} shows that the descent path for $P=0$ is the negative imaginary axis, and as a consequence the descent path transforms into such an axis when $P\to0$ discontinuously. In other words, there should exist a singularity when $P=0$ in the asymptotic expression for $\mu=0$, which indeed is the case as Eq. \eqref{eq-lt12a} reveals. Regardless of $\mu$ is zero or not, the calculation of the survival amplitude for $P=0$ has been discussed exhaustively in the literature and we will not reproduce these calculations here. We encourage the reader to see \cite{ourpaperPRA} and references therein. The former mathematical considerations lead us to conclude that $\mu=0$ is not an admissible solution {\it from a physical point of view} because it is not possible to recover the nonrelativistic 
case when $P=0$ because there is no term in the expansion which does not depend on $P$. {In \cite{urbanowski_rel, shirokov2004}, the authors derive analytical expressions assuming $\mu = 0$ and eventually arrive at the same conclusion, that is, their results do not reproduce the nonrelativistic case i.e., $P \to 0$. On the other hand, by considering the case where $\mu \ne 0$, the survival amplitude reduces to the nonrelativistic case \eqref{eq-math13}. Moreover, the survival probability $P_{lt} = |A_{lt}|^2$,  
at large times for $\mu\neq0$, is given by
\begin{equation}\label{P_general}
P_{lt}(\tau)\sim\NB{Q(\mu)\Gamma(\al+1)}^2\EP{{1+\dfrac{P^2}{\mu^2}}\,}^{\al+1}\,\GP{\dfrac{2x_0}{\tau}}^{2(\al+1)}.
\end{equation}
If the momentum is such that $P\gg\mu$, the above expresion admits an additional simplification, namely,
\begin{equation}\label{Plarge}
P_{lt}(\tau)\sim\NB{Q(\mu)\Gamma(\al+1)}^2\,\GP{\dfrac{2x_0P}{\mu\tau}}^{2(\al+1)}.
\end{equation}
In literature, Eqs. \eqref{P_general} and \eqref{Plarge} are also obtained in \cite{GiraldiHindawi,GiraldiJPhysA}. The expression obtained there recovers the nonrelativistic limit as well as it shows a singularity in $\mu=0$. 
In addition, we provide (i) 
an explanation for the cases $\mu =0$ and $\mu \neq 0$ 
in Eqs (\ref{NElargeRel2}), 
(ii) the explanation for which the survival amplitude for large times shows a discontinuity when the limit $\mu\to0$ is taken and for the difference 
in the exponent of the power law, and (iii) how this pathology does not enable us to recover the correct nonrelativistic limit when $\mu\to0$, mathematically speaking. Finally, we note from the behaviour of Eq. (\ref{P_general}) that we cannot expect a critical Lorentz factor $\gamma$ where the relativistic effect is maximum as found earlier in case of exponential decay (see (\ref{gammac})).}

\subsection{Short time regime}\label{shortt}
In this section we shall study how the relativistic survival amplitude given 
by Eq. \eqref{eq-math7} behaves when $\dfrac{\tau}{2x_0}\ll1$. All we need to 
do is to expand Eq. \eqref{eq-math7}  
in a Taylor series around $\tau=0$:
{
\begin{equation}\label{eq-st0}
a(\tau,P)=\sum_{n=0}^{\infty}\frac{1}{n!}\GP{\frac{\tau}{2x_0}}^n\,\frac{d^n}{dk^n}\GC{\int_{\mu}^{\infty}{d\xi\,\lam(\xi)\,e^{-ik\sqrt{\xi^2+P^2}}}}_{k=0}.
\end{equation}
In order to compute the derivatives, we need to analyze two things:
\begin{enumerate}[i)]
\item The uniform convergence of the integral $\int_{\mu}^{\infty}{d\xi\,\lam(\xi)\,e^{-ik\sqrt{\xi^2+P^2}}}$ in a finite domain $t<t_0$, which is proved from the Weierstrass test for uniform convergence, that is, the integrand is 
smaller than the density of states in $t<t_0$, and its integral over $\xi\geq\mu$ exists because of \eqref{momenta_H}.
\item The uniform convergence of the integral $\int_{\mu}^{\infty}{d\xi\,\lam(\xi)\,\NP{\xi^2+P^2}^{n/2}\,e^{-ik\sqrt{\xi^2+P^2}}}$ in a finite domain $t<t_0$. Once again, we rely on the Weierstrass test for uniform convergence. Since, for $\xi\geq\mu$,
\begin{equation}
\MB{\lam(\xi)\,\NP{\xi^2+P^2}^{n/2}\,e^{-ik\sqrt{\xi^2+P^2}}\,}
\leq
\xi^n\lam(\xi)\,\GP{1+\frac{P^2}{\mu^2}}^{n/2},
\end{equation}
and since $\int_{\mu}^{\infty}d\xi\,\xi^n\lam(\xi)$ converges for all $n=0,1,2,\dotsc$ because of \eqref{momenta_H}, the integral
\[
\int_{\mu}^{\infty}{d\xi\,\lam(\xi)\,\NP{\xi^2+P^2}^{n/2}\,e^{-ik\sqrt{\xi^2+P^2}}}
\]
also converges uniformly. 
\end{enumerate}
The uniform convergence of these two integrals allows us to take the derivative under the integral sign. As a result, the Taylor expansion of the survival amplitude reads:}
\begin{equation}\label{eq-st1}
a(\tau,P)=\sum_{n=0}^{\infty}\frac{1}{n!}\GP{\frac{\tau}{2ix_0}}^n\,I_n,
\end{equation}
where $I_n$ is defined as
\begin{equation}\label{eq-st2}
I_n=\int_{\mu}^{\infty}{d\xi\,\lam(\xi)\,\NP{\xi^2+P^2}^{n/2}}.
\end{equation}
The condition (\ref{moments1}) guarantees the finiteness of the integral $I_n$ for all n.
The survival probability for small times is given by
\begin{multline}\label{eq-st3}
S(\tau,P)=|a(\tau,P)|^2=\sum_{n=0}^{\infty}\frac{(-1)^n}{(2n)!}\GP{\frac{\tau}{2x_0}}^{2n}\sum_{m=0}^{2n}
(-1)^m\binom{2n}{m}I_mI_{2n-m}
\\=1-\tau^2\,\frac{I_2-I_1^2}{4x_0^2}+\tau^4\,\frac{I_4-4I_1I_3+3I_2^2}{192x_0^4}+\cdots.
\end{multline}
The relativistic survival probability, as well as the nonrelativistic one, for small times follows a quadratic law, and the particular features of the survival probability in this regime are dictated by the properties of the 
integrals $I_n$. To begin with, we shall study how the small time behavior for large momenta is. Recall that large $P$ here means that $P\gg 1+\mu$ and 
$P\gg x_0$, and only $I_n$ when $n$ is odd requires an additional treatment in order to calculate the asymptotic expansion for larges values of $P$ via Watson's lemma 
\cite{Ablowitz}. To do so, we need the following identity: 
\begin{equation}\label{eq-st4}
\frac{1}{\sqrt{\xi^2+P^2}}
=\frac{1}{\sqrt{\pi}}\int_{0}^{\infty}{\frac{dx}{\sqrt{x}}\,e^{-x(\xi^2+P^2)}},
\end{equation}
which converges for all real values of $\xi$ and $P$, and converges absolutely and uniformly in $\xi>0$ and $P>0$. Hence, $I_n$ for an odd $n=2s+1$, $s=0,1,2,\dotsc$ can be rewritten as:{
\begin{align}
I_{2s+1}&=\frac{1}{\sqrt{\pi}}\int_{\mu}^{\infty}
{d\xi\,\NP{\xi^2+P^2}^{s+1}\lam(\xi)\,\int_{0}^{\infty}{\frac{dx}{\sqrt{x}}\,e^{-x(\xi^2+P^2)}}}
\notag\\&=
\frac{1}{\sqrt{\pi}}
\sum_{l=0}^{s+1}\binom{s+1}{l}P^{2l}
\int_{\mu}^{\infty}d\xi\,\xi^{2s+2-2l}\lam(\xi)\int_{0}^{\infty}{\frac{dx}{\sqrt{x}}\,e^{-x(\xi^2+P^2)}}
\notag\\&=
\frac{1}{\sqrt{\pi}}
\sum_{l=0}^{s+1}\binom{s+1}{l}P^{2l}
\int_{0}^{\infty}{\frac{dx}{\sqrt{x}}\,e^{-xP^2}}
\int_{\mu}^{\infty}
{d\xi\,\xi^{2s+2-2l}\lam(\xi)\,e^{-x\xi^2}}.\label{eq-st5}
\end{align}
The change of the order of integration is justified by the uniform convergence of the integral \eqref{eq-st4}. Now, the Watson's lemma requires expanding the integral in powers of $x$. Since the integral $\int_{\mu}^{\infty}{d\xi\,\xi^n\lam(\xi)e^{-x\xi^2}}$ where $n$ is a nonzero positive integer, converges uniformly for $x\geq0$ because from the Weierstrass test, $|\xi^n\lam(\xi)e^{-x\xi^2}|<\xi^n\lam(\xi)$ and the integral $\int_{\mu}^{\infty}{d\xi\,\xi^n\lam(\xi)}<\infty$ because of Eq. \eqref{momenta_H}, we can perform the expansion in power of $x$ of the integral \eqref{eq-st5} by taking derivatives inside such an integral. 
This procedure is justified from the uniform convergence of the integral analyzed below. As a consequence, we obtain that $I_{2s+1}$ for large $P$ is equal to:
\begin{align}
I_{2s+1}
&\sim
\frac{1}{\sqrt{\pi}}
\sum_{l=0}^{s+1}\binom{s+1}{l}P^{2l}
\sum_{v=0}^{\infty}\frac{(-1)^v}{v!}\,\frac{\Gamma\NP{v+\tfrac{1}{2}}}{P^{2v+1}}
\int_{\mu}^{\infty}{d\xi\,\lam(\xi)\,\xi^{2v+2s-2l+2}},
\notag\\&
=(2x_0)^{2s+1}
\sum_{l=0}^{s+1}\binom{s+1}{l}
\sum_{v=0}^{\infty}\frac{(-1)^v}{2^{2v}}\binom{2v}{v}
{\GP{\frac{P}{2x_0}}^{2l-2v-1}\,\tilde{\delta}_{2v+2s-2l+2}},\label{eq-st6}
\end{align}
where $\tilde{\delta}_n$ is defined in Eq. \eqref{momenta_H}.} On the other hand, the integrals $I_n$ for even $n=2s$, $s=0,1,2,\dotsc$ can be written as a polynomial in $P$, so that no special treatment is needed. As a result,
\begin{equation}\label{eq-st7}
I_{2s}
=\sum_{l=0}^{s}\binom{s}{l}\,P^{2l}\int_{\mu}^{\infty}{d\xi\,\lam(\xi)\,\xi^{2s-2l}}
=(2x_0)^{2s}\sum_{l=0}^{s}\binom{s}{l}\,\GP{\frac{P}{2x_0}}^{2l}\tilde{\delta}_{2s-2l}.
\end{equation}
{Now, let us compute the survival probability for small times and large momenta by taking the leading contribution for $I_n$ given by Eqs. \eqref{eq-st6} and \eqref{eq-st7}, namely, the terms containing the highest exponent in $P$. From these two equations it is straightforward to prove that $I_n\sim P^n$, $P\to\infty$, and once we substitute this result in the equation \eqref{eq-st3}, we obtain
\begin{equation}
S(\tau,P)\sim\sum_{n=0}^{\infty}\frac{(-1)^n}{(2n)!}\GP{\frac{\tau}{2x_0}}^{2n}\sum_{m=0}^{2n}
(-1)^m\binom{2n}{m}P^m P^{2n-m}=1,
\end{equation}
since the summation over $m$ is zero for $n>0$ and unity for $n=0$, 
result that can be inferred from the application of the binomial theorem on $(1-1)^{2n}$. This calculation shows that we need to consider more terms in the asymptotic expansion of the integrals \eqref{eq-st6} and \eqref{eq-st7} in order to see how the survival probability for small times behaves for large momenta. 
Henceforth, by writing the asymptotic expresions for $I_1$ to $I_4$ explicity:
\begin{align}
I_1\sim&P+\frac{2\tilde{\delta}_2 x_0^2}{P}-\frac{2\tilde{\delta}_4 x_0^4}{P^3}+\frac{4\tilde{\delta}_6
   x_0^6}{P^5}-\frac{10\tilde{\delta}_8 x_0^8}{P^7}+\frac{28\tilde{\delta}_{10}
   x_0^{10}}{P^9}-\frac{84\tilde{\delta}_{12} x_0^{12}}{P^{11}}+\dotsc,\\
I_2\sim&P^2+4x_0^2\tilde{\delta}_2,\\ 
I_3\sim&P^3+6\tilde{\delta}_2 P x_0^2+\frac{6\tilde{\delta}_4x_0^4}{P}-\frac{4\tilde{\delta}_6 x_0^6}{P^3}+\frac{6\tilde{\delta}_8x_0^8}{P^5}-\frac{12\tilde{\delta}_{10} x_0^{10}}{P^7}+\frac{28\tilde{\delta}_{12}
   x_0^{12}}{P^9}-\frac{72\tilde{\delta}_{14} x_0^{14}}{P^{11}}+\dotsc,\\
I_4\sim&P^4+8P^2x_0^2\tilde{\delta}_2+16\tilde{\delta}_4,
\end{align}
} and as a result, the survival probability for large $P$ takes the following form:
\begin{align}
&S(\tau,P)=
1-\tau^2\,\frac{I_2-I_1^2}{4x_0^2}+\tau^4\,\frac{I_4-4I_1I_3+3I_2^2}{192x_0^4}+\cdots
\notag\\&
=1
-\tau^2
\Biggl[
\NP{\dd_4-\dd_2^2}\GP{\frac{x_0}{P}}^2
+2\NP{\dd_2\dd_4-\dd_6^2}\GP{\frac{x_0}{P}}^4
-\NP{\dd_4^2+4\dd_2\dd_6-5\dd_8}\GP{\frac{x_0}{P}}^6
\notag\\&\hspace{2cm}
+2\NP{2\dd_4\dd_6+5\dd_2\dd_8-7\dd_{10}}\GP{\frac{x_0}{P}}^8
-2\NP{2\dd_6^2+5\dd_4\dd_8+14\dd_2\dd_{10}-21\dd_{12}}\GP{\frac{x_0}{P}}^{10}+\cdots
\Biggr]
\notag\\&\ +\tau^4
\Biggl[
\frac{3\dd_4^2-4\dd_2\dd_6+\dd_8}{12}\GP{\frac{x_0}{P}}^4
-\frac{2\dd_4\dd_6-3\dd_2\dd_8+\dd_{10}}{3}\GP{\frac{x_0}{P}}^6
+\frac{2\dd_6^2+9\dd_4\dd_8-18\dd_2\dd_{10}+7\dd_{12}}{6}\GP{\frac{x_0}{P}}^8
\notag\\&\hspace{2cm}
-\frac{4\NP{\dd_6\dd_8+3\dd_4\dd_{10}-7\dd_2\dd_{12}+3\dd_{14}}}{3}\GP{\frac{x_0}{P}}^{10}+\cdots
\Biggr]
+\cdots
.\label{eq-st9}
\end{align}
We draw the reader's attention to the subtle cancellations 
of the terms in $I_1$, $I_2$, $I_3$ $\dots$, which leave only terms of the type $1/P^n$ in 
(\ref{eq-st9}).
For large $P$, notice that the coefficient for the $\tau^2$ term is proportional to the uncertainty of the square of the non--relativistic Hamiltonian $H^2$ evaluated in the initial state, and hence it is positive. Finally, the relativistic survival probability for small times and for large momentum in terms of $t$, $p$, $\delta_2$ and $\delta_4$ takes the following form:
\begin{equation}\label{relshort}
S(\tau,P)=\, 1-\NP{{\delta}_4- {\delta}_2^2}\GP{\frac{t^2}{4p^2}} 
+ \cdots,
\end{equation}
implying that large momenta, $p$, slow down the decay for small times.  
Performing a similar expansion for the ``time dilation" expression for the 
survival probability $P_0(t/\gamma)$ in (\ref{specialrel}) and rewriting in 
the same notation as above (here $P_0(t/\gamma)$ and $S(\tau/\gamma,0)$ 
are both survival probabilities evaluated in the rest 
frame of the particle moving with velocity $v$, but at a time 
$t/\gamma$ instead of $t$),
\begin{equation}\label{timedilatedshort}
S(\tau/\gamma,0) = 1 -\NP {\delta_2 - \delta_1^2}\GP{ \frac{t^2}{\gamma^2}} 
+ \cdots.
\end{equation}
In the ultrarelativistic (UR) 
regime, $p = \gamma m_0 v = m_0 \sqrt{(\gamma^2 - 1)}  
\simeq m_0 \gamma$, and we can rewrite (\ref{relshort}) in a form similar to 
(\ref{timedilatedshort}), namely,  
\begin{equation}\label{ultrarelshort}
S^{UR}(\tau,P) \simeq 1-\biggl(\frac{{\delta}_4- {\delta}_2^2}{4m_0^2} \biggr ) 
\, \GP{ \frac{t^2}{\gamma^2}}\,  
\end{equation} 
and note that the energy uncertainty in (\ref{timedilatedshort}) 
gets replaced by a different factor in (\ref{ultrarelshort}).
Note that for the exponential part, the ultrarelativistic and the time 
dilation relations for survival probability are the same. However, due to a 
different ``energy uncertainty" factor in front of $t/\gamma$
in (\ref{timedilatedshort}) and (\ref{ultrarelshort}), this does not seem to be 
the case at short times. A critical gamma as in (\ref{gammac}) in the 
exponential case does not appear here. It is also clear from 
(\ref{ultrarelshort}) that the larger the value of $p$, the larger the survival 
(or non-decay) probability will be a quadratic law at short times. In other words, the particle becomes longer-lived.

\begin{table}[htb!]
\centering
\begin{tabular}{c|c|c}\hline
Process & $\delta_2-\delta_1^2$ (MeV$^2$) & $({\delta}_4- {\delta}_2^2)/4m_0^2$ (MeV$^2$)\\\hline
$\Delta^{++}$ & 5682.5                 & 6303.2\\\hline
     $\rho^0$ & 26282                  & 42080\\\hline
        $Z_0$ & 87245                  & 169040\\\hline
      $\mu^-$ & $1.2054\times10^{-14}$ & $2.3275\times10^{-14}$\\\hline
        $K^+$ & $7.9145\times10^{-12}$ & $1.3164\times10^{-11}$\\\hline
\end{tabular}
\caption{Values of the coefficients in the quadratic term in $t/\gamma$ given in the Eqs. \eqref{timedilatedshort} and \eqref{ultrarelshort} for some decay processes. The density of states assumed is given in the Eq. \eqref{dos_numbers_2}. In all of the calculations, $\alpha=\frac{1}{2}$ and $\beta=1$. Data for each process is available in \cite{pdg}.}\label{tab1}
\end{table}

\begin{figure}[htb!]
\centering
\includegraphics[scale=0.45]{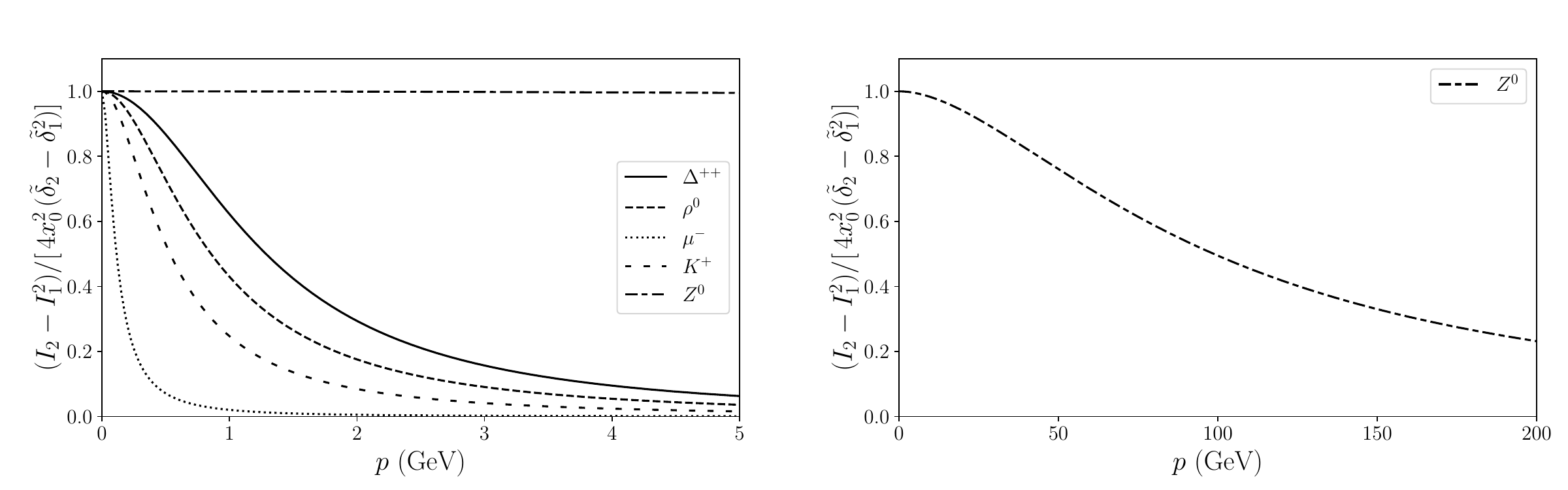}
\caption{On the left, plot of the ratio between the coefficients in the quadratic term in $t$ given in Eqs. \eqref{stimes_nr} and \eqref{eq-st3} for some decay processes. On the right, the plot corresponding to $Z^0$ for a larger range of $p$. The density of states assumed is given in Eq. \eqref{dos_numbers_2}. In all the calculations, $\alpha=\frac{1}{2}$ and $\beta=1$. Data for each process is available in \cite{pdg}.}\label{fig_uncert}
\end{figure}

In order to ilustrate these observations, in Table \ref{tab1}, we show 
the values of the coefficients in the quadratic term in $t/\gamma$ given in 
Eqs. \eqref{timedilatedshort} and \eqref{ultrarelshort} for some decay 
processes. In addition, in Figure \ref{fig_uncert} we compare the ratio 
between the coefficients in the quadratic term $t$ given in 
Eqs. \eqref{stimes_nr} and \eqref{eq-st3} for the same decay processes. 
These calculations were made by assuming a density of states times an exponential form factor given by
\begin{equation}\label{dos_numbers_1}
\omega(m)=N\,\frac{(m-\M)^\alpha}{(m-m_0)^2+\Gamma_0^2/4}\,e^{-\beta m/(m_0-\M)},
\end{equation}
or in terms of the parameters $x_0$ and $\mu$:
\begin{equation}\label{dos_numbers_2}
\lambda(\xi)=N'\,\frac{(\xi-\mu)^\alpha}{(\xi-1-\mu)^2+x_0^2}\,e^{-\beta\xi}.
\end{equation}
In these equations, $N$ and $N'$ are respectively normalization constants.

The coefficients for the term $\tau^2$ for the processes considered in general are a decreasing function of $p$, and therefore the survival probability of these processes in flight would have the small time non-exponential behaviour for a 
longer period of time than if these decays happened in a frame at rest. The lenght of such a period of time will depend of the values of $x_0$ and $\mu$ --recall that these two parameters determine when the momentum of the particle is considered large. Based on the values computed in Table \ref{tab21}, it is not surprising that $Z^0$ could be considered non--relativistic, and $\mu^-$ considered fully relativistic. These features are seen in Figure \ref{fig_uncert}, that is, the ratio of the coefficients of $t^2$ in flight and at rest, 
hardly changes for $Z^0$ but for $\mu^-$ it is almost zero when $p=1$ GeV. 
This decreasing indicates two things: i) the decay in flight is longer-lived in comparison to the same decay in a rest frame, and ii), the interval for which the survival probability is dominantly exponential is shrunk from the left. These properties are also visible when the critical times for small times are introduced and computed for the decays considered here. The latter is 
discussed in the next section.
%

\subsection{Critical times}\label{c_times}
In this section we shall study the critical times for both large and small times, and focus on how large momenta affect such times. 
We start with the critical time $\tau_{lt}$ associated with the transition 
from the exponential to the nonexponential regime. It is defined in the 
same way as it's nonrelativistic counterpart, that is, it will be the intersection of the exponential survival probability and nonexponential survival probability for large times:
\begin{equation}\label{eq_lt}
\NB{R}^2\,\exp{\NP{-\sigma\tau_{lt}}}=
\NB{Q(\mu)\Gamma(\al+1)}^2(2x_0)^{2\al+2}\EP{{1+\dfrac{P^2}{\mu^2}}\,}^{\al+1}\,{\tau_{lt}^{-(2\al+2)}}.
\end{equation}
By rearrenging this equation:
\begin{equation}
\NP{\sigma\tau_{lt}}^{\,2\al+2}\exp{\NP{-\sigma\tau_{lt}}}=\underbrace{\GB{\frac{Q(\mu)}{R}\Gamma(\al+1)}^2\EP{2x_0\,\sigma\,\sqrt{1+\dfrac{P^2}{\mu^2}}\,}^{2(\al+1)}}_{T},
\end{equation}
we can study if there exists a solution of this trascendental equation. Calling $j(w)=w^{\,2\al+2}\,e^{-w}$ with $w=\sigma\tau_{lt}$, we deduce that this function has one maximum at $w={2(\al+1)}$ and the function $j(w)$ takes in this maximum the value
\begin{equation}
j_{max}=j\NP{{2\al+2}}=\GC{\dfrac{2(\al+1)}{e}}^{2(\al+1)}.
\end{equation}
Since the function $j(\tau_{lt})$ is positive for $\tau_{lt}>0$ and it has one critical point (a maximum), we conclude that if $T>j_{max}$, there is no solution. If $T=j_{max}$, the solution will be the maximum itself. Finally, if $T<j_{max}$ there are two solutions and the critical time {\it is the largest of these 
two solutions}.

Let us write explicitly the ratio $\dfrac{T}{j_{max}}$:
\begin{equation}
\dfrac{T}{j_{max}}=\GB{\frac{Q(\mu)}{R}\Gamma(\al+1)}^2\GP{\frac{2ex_0}{\al+1}}^{2\al+2}\EP{{\sigma}\,\sqrt{1+\dfrac{P^2}{\mu^2}}\,}^{2(\al+1)}.
\end{equation}
For $P=0$, we can expect that the ratio will be less than one for narrow resonances, and for broad resonances there exists a value of $x_0$ for which there is 
no critical time. This fact has been tested exhaustively for nonrelativistic 
decay (see for instance, \cite{RamirezJPhysA} and references therein).

Once the relativity is included, we should recall that $\sigma$ decreases 
as $P$ increases, but it is compensate by the factor $\sqrt{1+\dfrac{P^2}{\mu^2}}$. Hence, we should not expect significant changes in $\sigma\tau_{lt}$. What it is interesting here is that for large $P$, the whole factor $\sigma\sqrt{1+\dfrac{P^2}{\mu^2}}$ is constant and equal to $\dfrac{1+\mu}{\mu}$. In other words, the quantity
\begin{equation}
w_0\equiv\lim_{P\to\infty}\sigma(x_0,P)\tau_{lt}(P)
\end{equation}
is constant and it is the largest solution of the equation
\begin{equation}\label{eq_w0}
w_0^{\,2\al+2}\exp{\NP{-w_0}}=\GB{\frac{Q(\mu)}{R}\Gamma(\al+1)}^2\EP{2x_0\,\frac{1+\mu}{\mu}\,}^{2(\al+1)}.
\end{equation}
Since $\sigma=O(P^{-1})$, we infer that $\tau_{lt}=O(P)$. Moreover, since $P\approx(1+\mu)\ga$ for large $P$, we have that
\begin{equation}\label{crittratio_a}
\tau_{lt}\NP{P\gg\NL{1+\mu,x_0}}\sim w_0\gamma.
\end{equation}

In the table \ref{tab2} we present calculations of some critical times 
at large times for a range of momenta for some decay processes, as well as the parameter $w_0$.
\begin{table}[htb!]
\centering
\begin{tabular}{c|c|c|c|c|c|c}\cline{3-7}
\multicolumn{2}{c}{} & \multicolumn{5}{c}{Critical time $\tau_{lt}$}\\\hline
Process & $w_0=(\sigma\tau_{lc})|_{P\to\infty}$ & $p=0$ &  $p=0.5$ & $p=1$ & $p=5$ & $p=10$ \\\hline
$\Delta^{++}$ & 10.193 & 10.754 & 11.508 & 13.537 & 42.719 & 83.417 \\\hline
$\rho^0$      & 12.458 & 16.341 & 16.996 & 21.549 & 81.438 & 161.05\\\hline
$Z^0$         & 30.505 & 67.065 & 47.449 & 45.228 & 40.096 & 38.024\\\hline
$\mu^-$       & 203.88 & 220.11 & 986.43 & 1940.5 & 9650.2 & 19297\\\hline
$K^+$         & 195.75 & 200.45 & 279.99 & 442.92 & 1992.4 & 3970.1\\\hline
\end{tabular}
\caption{Critical times in the large time region for some decay processes 
over a range of momenta, which are computed by taking the largest solution of 
the equation \eqref{eq_lt}. The density of states assumed is given in Eq. \eqref{dos_numbers_2}. In all of the calculations, $\alpha=\frac{1}{2}$ and $\beta=1$. Every value of $p$ is given in 
GeV. Data for each process is available in \cite{pdg}.}\label{tab2}
\end{table}
Comparing every critical time at each momentum per decay process shows that the intermediate time where the decay is dominantly exponential is lengthened when the system is in flight, and the dilatation of the intermediate regime is noticeable for narrow resonances. However, the narrowness of the resonance is not a general criterion to determine how much the intermediate regime will be dilated compared to the one in the frame where the system is at rest. In spite of the fact that the decay of $Z^0$ is considered a narrow one, we must take into account whether for the momenta considered they are large in the sense we defined in the section \ref{intert}.
 
\begin{figure}[htb!]
\centering
\includegraphics[scale=0.6]{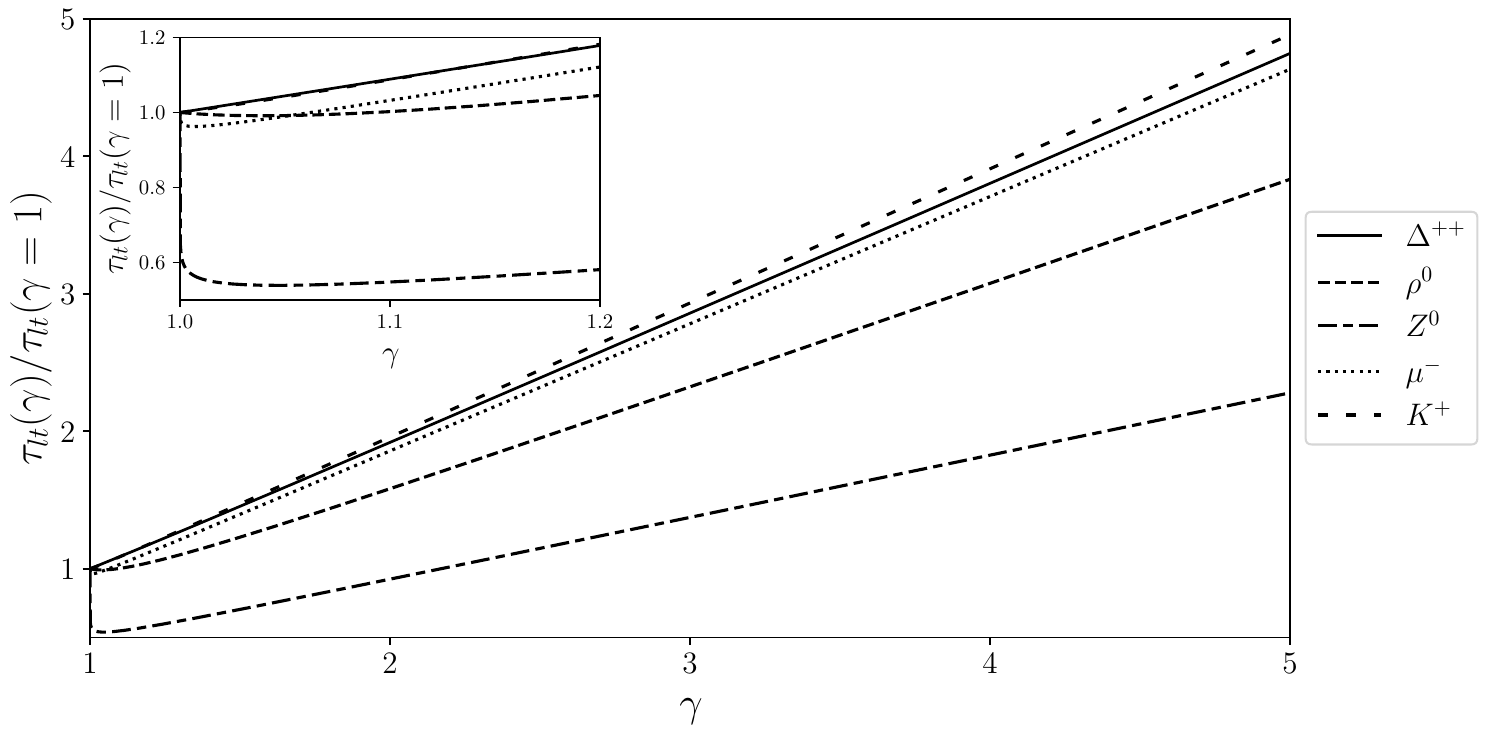}
\caption{Plot of $\tau_{lt}{\gamma}/\tau_{lt}(\gamma=1)$ vs. $\gamma$ for some decay process. Ratios were computed by taking the largest solution of 
the Eq. \eqref{eq_lt}. The density of states assumed is given in Eq. \eqref{dos_numbers_2}. In all of the calculations, $\alpha=\frac{1}{2}$ and $\beta=1$. Data for each process is available in \cite{pdg}.}\label{plot_ratio_ct}
\end{figure}

{By computing the associated values of $P$ for $Z^0$ (see table \ref{tab21}), we see that these values explain why the critical times change slowly, and therefore this process might be considered nonrelativistic in the sense defined here. On top of that, the critical times for $Z_0$ reveal another, apparently anomalous behavior, namely, the critical times decrese for an increasing $p$. Since the associated values of $P$ for this decay are not considered large momenta, one might inquire whether this behavior is commom or not. If we calculate how the critical time changes with respect to $\gamma$ from the Eq. \eqref{eq_lt}, it is possible to demonstrate that there might exist a value of $\gamma$, called $\gamma^*$, for which this derivative is zero such that the following equation must be 
satisfied:
\begin{equation}\label{eq_min_ct}
\sigma(\gamma^*,x_0)\tau(\gamma^*)=2(\al+1)\cfrac{\gamma^*+C\NC{2\sigma(\gamma^*,x_0)^2-1}}{{\gamma^*}^{\,2}-1+\GP{\cfrac{\mu}{1+\mu}}^2} \, ,
\end{equation}
and in addition, the second derivative of the critical time with respect to $\gamma$ is positive when it is evaluated at $\gamma=\gamma^*$, that is, there exists a value of $\gamma$ for which the critical time is minimum provided that the Eq. \eqref{eq_min_ct} has a solution. Here, $C=\NP{\tfrac{x_0}{1+\mu}}^2$ was defined in the Appendix \ref{app3}, and $\sigma$ is given in terms of $\gamma$ in 
Eq. \eqref{sig_gam}. Numerical calculations over the processes considered here 
validate this result and it suggests a rule of thumb for which Eq. \eqref{eq_min_ct} has solution, namely, the critical time in the rest frame is lower than the r.h.s of Eq. \eqref{eq_min_ct} evaluated at $\gamma=1$, that is,
\begin{equation}\label{ineq}
\tau_{lt}(\gamma=1)<2(\al+1)(\mu+1)^2\frac{C+1}{\mu^2}.
\end{equation}
This condition is satisfied for $\rho^0$, $Z^0$ and $\mu^-$, and the minima occur respectively at $\gamma^*\approx1.039$, $1.046$ and $1.007$. Finally, Figure \ref{plot_ratio_ct} shows the ratio of the critical times for large times in flight and in the rest frame as a function of $\gamma$ for the process considered here. How should we interpret this plot? Notice that there are three processes whose curves are close to each other, namely, $\Delta^{++}$, $\mu^-$ and $K^+$. If we compare their $w_0$ values with their critical times in the rest frame (see the second and third colums in Table \ref{tab2}), there is not much difference and by computing the ratios between $w_0$ and $\tau_{lt}(\gamma=1)$, we see that they are close to one. This explains why the curves for these processes 
are close and they have the same slope when $\gamma$ is large --recall the Eq. \eqref{crittratio_a}. If we compute the same ratio for $Z^0$ and $\rho^0$, which are respectively equal to 0.45 and 0.76, it explains why their curves do not lie close to those for $\Delta^{++}$, $\mu^-$ and $K^+$. All of this strange behavior arises from Eq. \eqref{eq_w0}, and in particular from the term 
$\frac{x_0}{\mu}$. The processes $\Delta^{++}$, $\mu^-$ and $K^+$ are such that $\frac{x_0}{\mu}$ is negligible with respect to one, and since the only difference between the equation that satisfies the critical time at rest and the equation satisfied by $w_0$ is the term $\frac{1+\mu}{\mu}$, no major changes are expected. However, if the term $\frac{x_0}{\mu}$ is not negligible with respect to one, $w_0$ will differ with respect to $\tau_{lt}(\gamma=1)$. For $Z_0$, $\frac{x_0}{\mu}\sim1$ and for $\rho^0$, $\frac{x_0}{\mu}\sim0.2$, and as a consequence, the $w_0$ for $Z_0$ differs from its $\tau_{lt}(\gamma=1)$ the most among the processes considered here.} 

In connection with the critical time for the small time to the exponential regime, we must determine the frequency of oscillation of the survival probability in the first place. In \cite{ourpaperPRA} it was demonstrated that such a frequency has to be extracted from the ratio between the exponential and nonexponential survival amplitude. Since we decompose the relativistic survival amplitude as a sum of an exponential and a nonexponential component, the corresponding survival probability is split in the same way as in the nonrelativistic frame; henceforth we define the frequency of oscillation for the relativistic survival probability in the same way as in the nonrelativistic formalism. From the exponential survival probability, we have a oscillating term given by $\exp{\NP{-i\Omega \tau}}$, and from the nonexponential survival amplitude, we see that there exists a whole phase factor $\exp{\NP{-\frac{i\tau}{2x_0}\,\sqrt{P^2+\mu^2}}}$. Hence, the oscillation of the survival amplitude will be determined by 
\[
\exp{\EC{i\tau\GP{\frac{1}{2x_0}\sqrt{P^2+\mu^2}-\Omega}}},
\]
and therefore the frequency of oscillation of the relativistic survival probability $\nu$ is given by
\begin{equation}
\nu\equiv\Omega-\frac{1}{2x_0}\sqrt{P^2+\mu^2}
\end{equation}
which reduces to the nonrelativistic case when $P=0$, that is, $\nu=\dfrac{1}{2x_0}$. On the other hand, this frequency for large momenta approximates as
\begin{equation}
\nu\sim\frac{1+2\mu-x_0^2}{4x_0P}+\cdots.
\end{equation}
In other words, $\nu=O(P^{-1})$ and as a consequence the frequency of oscillation decreases when $P$ is large. This result shows once more the slowness of the decay of an unstable system for large momenta as compared to the one at rest in 
this region. 

\begin{figure}[htb!]
\centering
\includegraphics[scale=0.6]{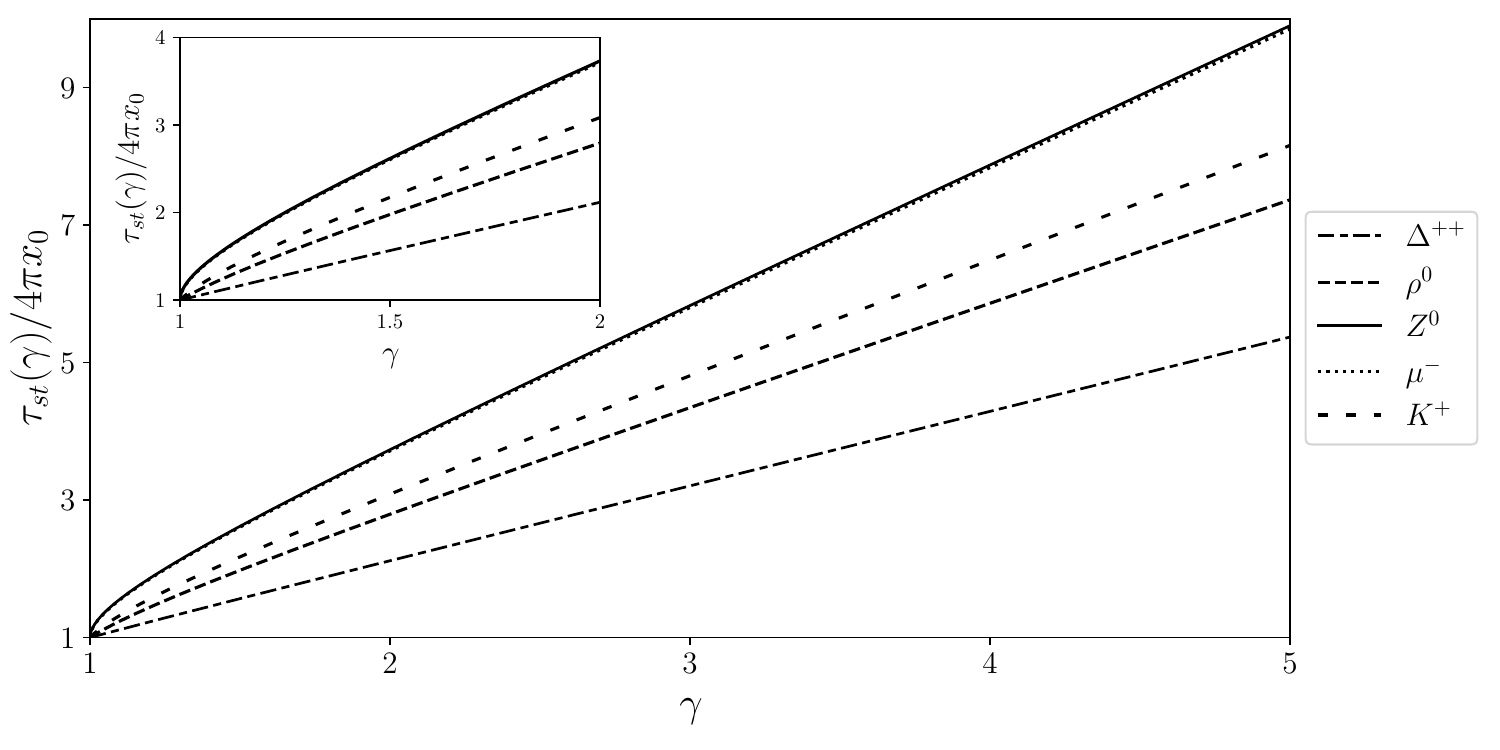}
\caption{Plot of $\tau_{st}{(\gamma)}/4\pi x_0$ vs. $\gamma$ for some decay process. Ratios were computed from the Eq. \eqref{crittshort1}. The density of states assumed is given in Eq. \eqref{dos_numbers_2}. In all of the calculations, $\alpha=\frac{1}{2}$ and $\beta=1$.}\label{ratio_st}
\end{figure}

Likewise, we define the critical time for small times as the time that it 
takes the unstable system to reach its first oscillation. Calling this time $\tau_{st}$, it is equal to
\begin{equation}
\tau_{st}\equiv\frac{2\pi}{\nu}=\frac{4\pi x_0}{2x_0\Omega-\sqrt{P^2+\mu^2}}.
\end{equation}
Notice that $4\pi x_0$ is nothing but the critical time in the nonrelativistic 
case \cite{ourpaperPRA}. Hence, it is convenient to write the above equation
as follows:
\begin{equation}\label{crittshort1}
\frac{\tau_{st}}{4\pi x_0}=\frac{1}{2x_0\Omega-\sqrt{P^2+\mu^2}}.
\end{equation}
If we expand $\Omega$ as an asymptotic series around $P=\infty$, that is,
\begin{equation}
\Omega=\frac{P}{2x_0}+\frac{1-x_0^2+2\mu+\mu^2}{4Px_0}+O\NP{P^{-2}},
\end{equation}
we obtain that the critical time for small times for large momenta is given by:
\begin{equation}\label{crittshort2}
\frac{\tau_{st}}{4\pi x_0}\sim\frac{2P}{1+2\mu-x_0^2}+\cdots=\frac{2(1+\mu)}{1+2\mu-x_0^2}\gamma+\cdots.
\end{equation}
In Table \ref{tab3} we calculate the ratio between the relativistic and nonrelativistic critical time for small times given by Eq. \eqref{crittshort1} at different momenta for some decay processes. Likewise, in Figure \ref{ratio_st} we plot the same ratio in terms of $\ga$.

\begin{table}[htb!]
\centering
\begin{tabular}{c||c|c|c|c|c}\cline{2-6}
\multicolumn{1}{c}{} & \multicolumn{5}{c}{$\tau_{st}/4\pi x_0$}\\\hline
Process       & $p=0$ &  $p=0.5$ & $p=1$ & $p=5$ & $p=10$\\\hline
$\Delta^{++}$ & 1 & 1.0914 & 1.3283 & 4.4840 & 8.7991 \\\hline
$\rho^0$      & 1 & 1.4209 & 2.1962 & 9.6375 & 19.182 \\\hline
$Z^0$         & 1 & 1.0055 & 1.0110 & 1.0563 & 1.1157\\\hline
$\mu^-$       & 1 & 9.5228 & 18.890 & 94.198 & 188.38\\\hline
$K^+$         & 1 & 2.0251 & 3.5386 & 16.728 & 33.391\\\hline
\end{tabular}
\caption{Ratio between  the relativistic and nonrelativistic critical time for small times given by Eq. \eqref{crittshort1}. Every value of $p$ is given in 
GeV. Data for each process is available in \cite{pdg}.}\label{tab3}
\end{table}

Comparing every critical time at each momentum per decay process shows that the starting of the exponential regime in general is delayed when the system is in flight, as observed in the analysis of the relativistic survival probability for short times. In addition, the delay is critical for processes where both $x_0$ and $\mu$ are closed to zero, and under these conditions, $\tau_{st}{(\gamma)}/4\pi x_0\sim 2\gamma$ when $\gamma\gg1$ --this is ilustrated in the figure \ref{ratio_st} for $Z^0$ and $\mu^-$. Unlike the critical time for large times, the one for small times is a increasing function of $P$ (or $\gamma$) and accordingly there is no value for $\gamma$ for which this critical time has a maximum or a minimum.

We would like to close this section about the critical times by summarizing 
briefly the results found here. In general, a decay in flight compared with 
the same decay in a rest frame is such that the starting of the exponential 
regime will be delayed because the critical time for small times is larger 
in flight with respect to the one in the rest frame. Moreover, the starting 
of the large time regime might go forward or backward with respect to the same 
decay in a rest frame. The latter is due to the fact that depending 
on the resonance parameters, $x_0$ and $\mu$ there exists a critical value of 
$\gamma$ which decides the direction of shift at a given momentum.

%
\section{Summary and Outlook}\label{summ}
In fundamental physics, the spontaneous decay of metastable states 
is an intrinsic prediction
of quantum mechanics coupled to Special Relativity where the
fundamental processes are restricted only by conservation laws like
energy-momentum conservation, charge conservation and also lepton and
baryon number conservation. This makes a spontaneous decay possible. 
One encounters the phenomenon of decay in
many branches of physics with typical examples being, 
$Z^0 \to e^+ e^-$ in particle physics, $^{212}$Po $\to \, ^{208}$Pb + 
$\alpha$ in nuclear physics and $\Delta \to \pi N$ in hadron 
physics, just to mention a few. However, 
we can find it also in atomic and molecular physics, along with
investigations of the nonexponential decay law \cite{wilkinson,andersson,
pepe}. This explains
its importance in natural sciences. Motivated by the recent results
regarding the time evolution of an unstable state in flight, 
we have reconsidered the topic
specializing on three different time regimes: the small time survival
probablity of the form $S(t,p)=1-b(p)t^2 +...$, its equivalent for
intermediate times of the form $S(t,p)=\exp(-a(p)t)$ and the large time
behavior where the survival probablity is a power 
law $S(t,p)\propto p^{2(\alpha + 1)}/t^{2(\alpha + 1)}$, where $\alpha$ 
is the exponent in the threshold factor.
In doing so we employed the steepest descent method to calculate the
integrals involved. In the exponential region, we find a critical 
Lorentz factor $\gamma$, 
i.e., a $\gamma_c$ for
which the deviation from $\exp(-\Gamma t/\gamma)$ is the largest. It
corresponds to a velocity $v=0.63$c. This a purely quantum mechanical 
effect since, replacing $e^{-\Gamma t}$ at rest by $e^{-\Gamma t/\gamma}$ 
in flight, one is using relativity based on classical principles. 
For small times $b(p)$ includes a quantum mechanical expression for the energy
uncertainty. This is also the case for the decay at rest, but
surprisingly both uncertainties differ when the momentum $p$ is large.

The critical times of transition from the 
nonexponential short time regime to the exponential one ($\tau_{st}$) 
and the exponential 
to the large time power law regime ($\tau_{lt}$) 
are also presented for decays in flight. 
These times show an involved dependence on the resonance mass, $m_0$ and 
its width, the threshold energy, $m_{th}$ and momentum, $p$, 
of the decaying particle but reduce to rather simple analytical formulae 
for very large values of the parameter $P = p/(m_0 - m_{th})$.
An interesting feature here is that there exists a critical value of the 
Lorentz factor, $\gamma$, which decides if the nonexponential region at 
large times sets in at earlier or later times as compared to the one for
a particle at rest.    
Numerical calculations of some realistic resonances display interesting 
features. The most peculiar behaviour is that of the heavy $Z^0$ boson due
its large mass as compared to the other hadron and lepton resonances studied
here.

In this article we have exploited a relativistic and quantum mechanical (RQM) 
approach for studying different charateristics of the survival probability of 
unstable particles in flight. As 
mentioned in Section (\ref{section2}), 
this theory is not unanimously accepted and
there exist other results in literature. 
In view of this fact, analytical expressions for the 
non-relativistic (NR), relativistic (R) and ultrarelativistic (UR) 
regimes are provided for each of the three regions (the short time 
nonexponential, intermediate time exponential and large time nonexponential) 
of the decay law $S(t,p)$ in flight  
and compared with the time dilation formula, $P_0(t/\gamma)$. The UR limits
of the survival probabilities in the exponential and large time regions  
coincide exactly with $P_0(t/\gamma)$. In the short time region, the UR limit
of $S(t,p)$ has a similar form as that of $P_0(t\gamma)$ but with a different 
energy uncertainty. 

Given the differences in approaches, one can say that 
the fate of the decay law is
still not decided and this work means to provide some 
insight through the numerical calculations as well as the 
analytical results derived here. 
In times of
theoretical uncertainty, an experiment  
could be pivotal to decide the
issue. But such an undertaking would most likey require a very high
precision. In this context we note that it has been observed \cite{rindler} 
that up to a certain precision, the decay law $e^{-\Gamma t/\gamma}$ is also
observed for accelerated particles. This raises the question if the 
modification presented here would also be applicable for accelerated 
particles. If so, this may give hope for new experimental ideas.
\begin{acknowledgments}
N.G.K. thanks the Faculty of Science, Universidad de Los Andes, Colombia, for financial support through Grant No. INV-2023-162-2841.
\end{acknowledgments}
\appendix
\section{Properties of $\sqrt{(\zeta_0+\mu)^2+P^2}$}\label{app2}
In this section we shall study some properties of the complex number
\begin{equation}\label{app2-1}
\zeta=\sqrt{(\zeta_0+\mu)^2+P^2}\equiv x_0(2\Omega-i\sigma),
\end{equation}
with $\zeta_0=1-ix_0$, and $x_0>0$. To begin, let us compute the real and imaginary part of $\zeta^2$:
\begin{align*}
\Re{\zeta^2}&=P^2+(1+\mu)^2-x_0^2,\\
\Im{\zeta^2}&=-2x_0(1+\mu).
\end{align*}
We infer that the imaginary part is always negative, and so is the imaginary part of $\zeta$. From the definition of $\sigma$, it must be positive.

In addition, let us compute explicity the real and imaginary part of $\zeta$. To do so, take a complex number $z=re^{\,i\theta}$, and its square root is $r^{1/2}e^{\pm i\theta/2}$. But:
\[
\sqrt{z}=r^{1/2}\GP{\cos{\frac{\theta}{2}\pm i\sin{\frac{\theta}{2}}}}=
\sqrt{\frac{r+r\cos{\theta}}{2}}\pm i\,\sqrt{\frac{r-r\cos{\theta}}{2}}=
\sqrt{\frac{|z|+\Re{z}}{2}}\pm i\,\sqrt{\frac{|z|-\Re{z}}{2}}.
\]
Hence, we need only to compute the magnitude and the real part of the complex number whose square root we wish to calculate and take the correct sign for the imaginary part. In our particular case, we need only to make an additional calculation, that is,
\[
\NB{\zeta^2}^2=\NC{(P+x_0)^2+(1+\mu)^2}\NC{(P-x_0)^2+(1+\mu)^2},
\]
and from the Eq. \eqref{app2-1} $\sigma$ and $\Omega$ are easily calculated. As a result,
\begin{align}
\Omega&=
\frac{\Re{\zeta}}{2x_0}=\frac{1}{2x_0}\,\sqrt{\frac{|\zeta^2|+\Re{\zeta^2}}{2}}
\notag\\&=
\frac{1}{2\sqrt{2}x_0}\GC{\sqrt{\NC{(P+x_0)^2+(1+\mu)^2}\NC{(P-x_0)^2+(1+\mu)^2}}+P^2+(1+\mu)^2-x_0^2}^{1/2},\label{app2-2}
\\
\sigma&=-\frac{1}{x_0}\Im{\zeta}=\frac{1}{x_0}\sqrt{\frac{|\zeta^2|-\Re{\zeta^2}}{2}}
\notag\\&=
\frac{1}{\sqrt{2}x_0}\GC{\sqrt{\NC{(P+x_0)^2+(1+\mu)^2}\NC{(P-x_0)^2+(1+\mu)^2}}-P^2-(1+\mu)^2+x_0^2}^{1/2}.\label{app2-3}
\end{align}
Finally, since $\Im{\zeta^2}=2\Re{\zeta}\Im{\zeta}=-4x_0^2\sigma\Omega$, we have the following relation:
\begin{equation}
\sigma\,\Omega=\frac{1+\mu}{2x_0},
\end{equation}
in other words, $\sigma$ and $\Omega$ are not independent.
\section{Properties of the function $f(\gamma,x_0)$}\label{app3}
Before starting the study of the function $(\gamma,x_0)$, we need to rewrite $\sigma$ in terms of $\gamma$. From the definition of the relativistic momentum, we have that $p=m_0v\ga$, and since $\ga^2(1-v^2)=1$, $p^2=m_0^2(\ga^2-1)$. On the other hand, from the definition of the $P$ parameter, we get 
\[
P^2=\frac{p^2}{(m_0-M)^2}=\GP{\frac{m_0}{m_0-M}}^2(\ga^2-1)=(1+\mu)^2(\ga^2-1)
\]
and substituting this in $\sigma$ and after some manipulations, we obtain the following relation:
\begin{equation}\label{sig_gam}
\sigma^2(\gamma,x_0)=\frac{1}{2}\GP{\frac{1+\mu}{x_0}}^2\,
\EL{\sqrt{\GC{\ga^2-\GP{\frac{x_0}{1+\mu}}^2}^2+4\GP{\frac{x_0}{1+\mu}}^2}-\ga^2+\GP{\frac{x_0}{1+\mu}}^2}\, .
\end{equation}
Now, let us introduce the following variables in order to simplify the calculations we need to perform in a while, namely,
\begin{align}
y&=\ga^2-\GP{\frac{x_0}{1+\mu}}^2,\\
C&=\GP{\frac{x_0}{1+\mu}}^2.
\end{align}
If $0<x_0<1$, $C$ must be in the interval $0<C<1$. Hence, $\sigma$ transforms as
\begin{equation}
\sigma^2(y,C)=\frac{\sqrt{y^2+4C}-y}{2C}=\frac{2}{\sqrt{y^2+4C}+y}.
\end{equation}
Now we are in a suitable position to study the function $f(\gamma)$:
\begin{enumerate}[i)]
\item It turns out that for constant $x_0$, $f(\infty,x_0)=0$. 

This can be proved as follows: if $P\to\infty$, $\gamma\to\infty$ and thus $y\to\infty$, and from the definition of $\sigma$, 
\begin{equation}
\lim_{y\to\infty}\sigma(y,C)=\lim_{y\to\infty}\sqrt{\frac{2}{\sqrt{y^2+4C}+y}}=0,
\end{equation}
and as a consequence of this limit, $f(\infty,x_0)=0$. This result implies that, for constant $x_0$,
\begin{equation}
\lim_{\ga\to\infty}\frac{P_e(\tau,P)}{p_e(\tau,\ga)}=1.
\end{equation}
\item Now, we would like to compute the critical points of $f(\gamma)$ for constant $x_0$:
\begin{equation}
\frac{\partial f(\gamma,x_0)}{\partial\ga}=\frac{\partial\sigma(y(\gamma,C),C)}{\partial\ga}+\frac{1}{\ga^2}=\frac{\partial\sigma(y,C)}{\partial y}\frac{\partial y(\gamma,C)}{\partial\ga}+\frac{1}{\ga^2}=\frac{1}{\ga^2}-\frac{\ga\sigma}{\sqrt{y^2+4C}}
\end{equation}
and ${\partial f(\gamma,x_0)}/{\partial\ga}=0$ implies that
\begin{equation}
\sigma=\frac{\sqrt{y^2+4C}}{\ga^3}.
\end{equation}
From the definition of $\sigma$ and $\gamma$:
\begin{equation}
\sigma^2=\frac{\sqrt{y^2+4C}-y}{2C}=\frac{y^2+4C}{\ga^6}=\frac{y^2+4C}{(y+C)^3},
\end{equation}
and from here we can obtain the equivalent relation
\begin{equation}
\frac{1}{\sigma^2}=\frac{\sqrt{y^2+4C}+y}{2}=\frac{(y+C)^3}{y^2+4C},
\end{equation}
which lets us get rid of the term $\sqrt{y^2+4C}$ by substracting both equations. Hence,
\begin{equation}
y=\frac{(y+C)^3}{y^2+4C}-C\frac{y^2+4C}{(y+C)^3}.
\end{equation}
This equation leads to a quintic equation in $Z=y+C=\gamma^2$, namely,
\begin{equation}
3Z^5-(3C+5)Z^4+C(C+8)Z^3-2C(3C+4)Z^2+4C^2(C+4)Z-C^2(C+4)^2=0.
\end{equation}
Despite the fact that this equation does not have an analytic solution, we know that this equation has at least one real solution, and it is positive because of the Descartes'\ sign root (we can prove this by making $Z\to-Z$ and see that there is no variation in the signs of the coefficients of the equation). In addition, we can make an approximation for small $C$. In this case we neglect all powers of $C$ and the equation reduces to
\begin{equation}
Z^4(3Z-5)=0,
\end{equation}
Hence, a solution of this equation for small $C$ is $\gamma=\sqrt{\dfrac{5}{3}}+O(C)=1.2909\dotsc+O(C)$.

Moreover, it is possible to improve this solution if we assume that the root can be written in the form
\begin{equation}
Z=\frac{5}{3}+a_1C+a_2C^2+a_3C^3+a_4C^4+\cdots.
\end{equation}
If we substitute this form into the quintic equation, we can find the coefficients iteratively, and therefore the solution can be written in the following form:
\begin{equation}
Z=\ga^2=\frac{5}{3}+\frac{9
   C}{25}-\frac{31 C^2}{3125}+\frac{267 C^3}{390625}-\frac{573
   C^4}{9765625}+\frac{33642 C^5}{6103515625}+\cdots.
\end{equation}
or in terms of $x_0$:
\begin{multline}
\ga^2=\frac{5}{3}+\frac{9}{25}\GP{\frac{x_0}{1+\mu}}^2
-\frac{31}{3125}\GP{\frac{x_0}{1+\mu}}^4\\
+\frac{267}{390625}\GP{\frac{x_0}{1+\mu}}^6-\frac{573}{9765625}\GP{\frac{x_0}{1+\mu}}^8+\frac{33642}{6103515625}\GP{\frac{x_0}{1+\mu}}^{10}+\cdots.
\end{multline}
It turns out that this quintic equation has one real positive solution given by the above power series, and this value is a maximum of the function $f(\gamma)$ for constant $x_0$. Hence we conclude that the ratio $\dfrac{P_e(\tau,P)}{p_e(\tau,\ga)}$ has a minimum when $\ga^2$ is given by this power series. Notice that, for narrow resonances, this minimum is such that $\ga^2\approx\dfrac{5}{3}$.
\item Now let us study how $f(\gamma)$ behaves with respect to $x_0$ for constant $\gamma$, and this behavior reduces to analyzing how $\sigma$ changes with respect to $C$. Firstly, if $x_0$ tends to zero, $C\to0$ and it is easy to prove that $\sigma\to\dfrac{1}{\ga}$, and thus $f(\gamma)\to0$. In other words,
\begin{equation}
\lim_{x_0\to0}\frac{P_e(\tau,P)}{p_e(\tau,\ga)}=1.
\end{equation}
Secondly, let us write $\sigma$ in terms of $\gamma$ and $C$:
\begin{equation}
\sigma^2=\frac{2}{\sqrt{y^2+4C}+y}=\frac{2}{\sqrt{(\gamma^2-C)^2+4C}+\gamma^2-C},
\end{equation}
and once we compute the derivative of $\sigma$ with respect to $C$, namely,
\begin{equation}
\frac{d\sigma}{dC}=\frac{\sigma}{2}\frac{1}{\sqrt{(\gamma^2-C)^2+4C}+\gamma^2-C}\frac{\sqrt{(\gamma^2-C)^2+4C}+\gamma^2-C-2}{\sqrt{(\gamma^2-C)^2+4C}}
=\frac{\sigma}{2}\frac{1-\sigma^2}{\sqrt{(\gamma^2-C)^2+4C}};
\end{equation}
thus $\sigma$ will have a critical point if $\sigma=1$, which is reached for the extreme narrow resonances (i.e., $x_0\to1$). On the other hand, since $\ga^2>1$, we can write the following chain of inequalities:
\begin{align}
&
\ga^2>1\Rightarrow\ga^2-C>1-C,
\notag\\&
\ga^2-C>1-C\Rightarrow(\ga^2-C)^2+4C>(1-C)^2+4C=(1+C)^2,
\notag\\&
\frac{2}{\sigma^2}=\sqrt{(\ga^2-C)^2+4C}+\ga^2-C>1+C+1-C=2C,\notag
\end{align}
and as a result, $\sigma^2<1$. In conclusion, $\sigma$ (and so does $f(\ga)$) is a monotonic increasing function for $C$ if $0<C<1$ and constant $\ga$. 

The consequence of this property is that the ratio $\dfrac{P_e(\tau,P)}{p_e(\tau,\ga)}$ decreases monotonically when $x_0$ goes from 0 to 1 for a constant $P$ (or constant $\ga$), or paraphrasing this statement, ${P_e(\tau,P)}$ deviates from ${p_e(\tau,\ga)}$ the most for broad resonances for constant $\ga$.
\end{enumerate}

\section{Analytical expresions for $I_1$ and $I_2$ for an exponential form factor}\label{app4}
{The aim of this section is to compute analytical expressions for the integrals
\begin{align}
I_1&=\int_{\mu}^{\infty}d\xi\,\lambda(\xi)\NP{\xi^2+P^2}^{1/2},\label{I1_ap}\\
I_2&=\int_{\mu}^{\infty}d\xi\,\lambda(\xi)\NP{\xi^2+P^2},\label{I2_ap}
\end{align}
if the density of states has the following form:
\begin{equation}\label{dos_ap}
\lambda(\xi)=\frac{(\xi-\mu)^\alpha}{(\xi-1-\mu)^2+x_0^2}\,e^{-\beta\xi},\quad\beta>0.
\end{equation}
Let us first compute $I_2$, for which we need the following expression:
\begin{align}
\cfrac{\xi^2+P^2}{\NP{\xi-\mu-1}^2+x_0^2}
&=
\cfrac{\NP{\xi-\mu-1}^2+2\NP{\mu+1}\NP{\xi-\mu-1}+P^2+\NP{\mu+1}^2}{\NP{\xi-\mu-1}^2+x_0^2}
\notag\\&=
1+\cfrac{P^2+\NP{\mu+1}^2-x_0^2+2\NP{\mu+1}\NP{\xi-\mu-1}}{\NP{\xi-\mu-1-ix_0}\NP{\xi-\mu-1+ix_0}}
\notag\\&=
1+\cfrac{P^2+\NP{\mu+1}^2-x_0^2+2ix_0\NP{\mu+1}}{2ix_0\NP{\xi-\mu-1-ix_0}}
-\cfrac{P^2-\NP{\mu+1}^2-x_0^2-2ix_0\NP{\mu+1}}{2ix_0\NP{\xi-\mu-1+ix_0}}
\notag\\&=
1-\frac{1}{x_0}\Im{\EC{\cfrac{P^2+(1+\mu-ix_0)^2}{\xi-\mu-1+ix_0}}}.\label{eq-st7a}
\end{align}
Substituting Eqs. \eqref{eq-st7a} and \eqref{dos_ap} in Eq. \eqref{I2_ap}, we have:
\begin{align}
I_2
&=
\int_{\mu}^{\infty}{d\xi\,\NP{\xi-\mu}^\al\,e^{-\beta\xi}}
-\frac{1}{x_0}\Im{\EC{\NC{P^2+(1+\mu-ix_0)^2}\int_{\mu}^{\infty}{d\xi\,\cfrac{\NP{\xi-\mu}^\al}{\xi-(1+\mu-ix_0)}\,e^{-\beta\xi}}}}
\notag\\&=
{\int_{\mu}^{\infty}{d\xi\,\NP{\xi-\mu}^\al\,e^{-\beta\xi}}
-\frac{1}{x_0}\Im{\EC{\NC{P^2+(\xi_0+\mu)^2}\int_{\mu}^{\infty}{d\xi\,\cfrac{\NP{\xi-\mu}^\al}{\xi-\mu-\xi_0}\,e^{-\beta\xi}}}}},\label{eq-st9a}
\end{align}
where $\xi_0=1-ix_0$. Making the change of variables $\xi\to\xi-\mu$:
\begin{align}
I_2
&=
e^{-\beta\mu}\EC{\int_{0}^{\infty}{d\xi\,{\xi}^\al\,e^{-\beta\xi}}
-\frac{1}{x_0}\Im{\GP{\NC{P^2-(\xi_0+\mu)^2}\int_{0}^{\infty}{d\xi\,\cfrac{{\xi}^\al}{\xi-\xi_0}\,e^{-\beta\xi}}}}}
\notag\\&=
\Gamma{\NP{\al+1}}e^{-\beta\mu}
\GL{\frac{1}{\beta^{\al+1}}
-\frac{1}{x_0}\Im{\MC{\NC{P^2-(\xi_0+\mu)^2}(-\xi_0)^\al\,e^{-\beta\xi_0}\Gamma{\NP{-\al,-\beta\xi_0}}}}},\label{eq-st18}
\end{align}
where $\Ga(\nu,z)$ is the upper incomplete gamma function. The integral inside the imaginary part of the function is obtained from the integral \cite{Erderly}
\[
\int_{0}^{\infty}\frac{x^{\al}}{x+\sigma}\,e^{-sx}\,dx=\Ga(\al+1)\sigma^\al\,e^{s\sigma}\Ga(-\al,s\sigma),\quad \Re{s}>0,\,|\Arg{\sigma}|<\pi.
\]
Regarding $I_1$: substituting Eq. \eqref{dos_ap} in Eq. \eqref{I1_ap}, we have: 
\begin{equation}\label{eq-st10}
I_1=
\int_{\mu}^{\infty}{\frac{d\xi}{\sqrt{\xi^2+P^2}}\,\cfrac{\NP{\xi-\mu}^\al\NP{\xi^2+P^2}}{\NP{\xi-\mu-1}^2+x_0^2}\,e^{-\beta\xi}}.
\end{equation}
Now, the factor $\NP{\xi^2+P^2}^{-1/2}$ can be rewritten as the result of the following integral:
\begin{equation}
\frac{1}{\sqrt{\xi^2+P^2}}
=\frac{1}{\sqrt{\pi}}\int_{0}^{\infty}{\frac{dx}{\sqrt{x}}\,e^{-x(\xi^2+P^2)}},\label{eq-st11}
\end{equation}
which converges for all real values of $\xi$ and $P$. Thus, the integral $I_1$ transforms as:
\begin{align}\label{I1_ap2}
I_1&=\frac{1}{\sqrt{\pi}}\int_{\mu}^{\infty}d\xi\,\int_{0}^{\infty}dx\,
\cfrac{\NP{\xi-\mu}^\al\NP{\xi^2+P^2}}{\NP{\xi-\mu-1}^2+x_0^2}\,e^{-\beta\xi}\,
\frac{1}{\sqrt{x}}\,e^{-x(\xi^2+P^2)}.
\end{align}
In order to justify the change of the order of integration via Weierstrass test, we need to study whether the integrand is bounded by a function that does not depend on $\xi$. If we write the integrand as
\begin{equation}
\EB{\cfrac{\NP{\xi-\mu}^\al\NP{\xi^2+P^2}}{\NP{\xi-\mu-1}^2+x_0^2}\,e^{-\beta\xi}\,
\frac{1}{\sqrt{x}}\,e^{-x(\xi^2+P^2)}}
\leq
\frac{e^{-xP^2}}{\sqrt{x}}\EB{\cfrac{\xi^2+P^2}{\NP{\xi-\mu-1}^2+x_0^2}}
\cdot\NB{\NP{\xi-\mu}^\al\,e^{-\beta\xi}},
\end{equation}
thus in $\xi>\mu$ we have the following inequalities:
\begin{align}
&\EB{\cfrac{\xi^2+P^2}{\NP{\xi-\mu-1}^2+x_0^2}}=
\EB{1-\frac{1}{x_0}\Im{\EC{\cfrac{P^2+(\mu+\xi_0)^2}{\xi-\mu-\xi_0}}}}
\leq
1+\EB{\frac{P^2+(\mu+\xi_0)^2}{\xi_0}},\label{ineq1}\\
&\NB{\NP{\xi-\mu}^\al\,e^{-\beta\xi}}\leq
\max_{\xi>\mu}\NC{\NP{\xi-\mu}^\al\,e^{-\beta\xi}}=\GP{\frac{\al}{\beta}}^\al\,e^{-\beta\mu-\al}.
\end{align}
As a consequence, the integrand satisfies the inequality in $\xi>\mu$ and $x>0$
\begin{equation}
\EB{\cfrac{\NP{\xi-\mu}^\al\NP{\xi^2+P^2}}{\NP{\xi-\mu-1}^2+x_0^2}\,e^{-\beta\xi}\,
\frac{1}{\sqrt{x}}\,e^{-x(\xi^2+P^2)}}
\leq
\EP{1+\EB{\frac{P^2+(\mu+\xi_0)^2}{\xi_0}}}\,\GP{\frac{\al}{\beta}}^\al\,e^{-\beta\mu-\al}\,\frac{e^{-xP^2}}{\sqrt{x}},
\end{equation}
and since the integral $\int_{0}^{\infty}dx\,x^{-1/2}\,{e^{-xP^2}}$ converges for $P>0$, the integral in $x$ inside the Eq. \eqref{I1_ap2} converges uniformly for $\xi>\mu$. Henceforth the order of integration can be changed thanks to the Weiertrass test, and as a result,
\begin{equation}\label{eq-st12}
I_1\sqrt{\pi}
=\int_{0}^{\infty}{\frac{dx}{\sqrt{x}}\,e^{-P^2x}
\int_{\mu}^{\infty}{{d\xi}\,\cfrac{\NP{\xi-\mu}^\al\NP{\xi^2+P^2}}{\NP{\xi-\mu-1}^2+x_0^2}\,e^{-\beta\xi}\,e^{-x\xi^2}}}.
\end{equation}
Next, the term $e^{-x\xi^2}$ can be expanded in terms of Hermite polynomials, namely \cite{Lebedev},
\begin{equation}\label{eq-st19}
e^{-a^2\xi^2}=\sum_{n=0}^{\infty}\frac{(-1)^n\,a^{2n}}{2^{2n}\,n!\,(1+a^2)^{n+1/2}}\,H_{2n}(\xi),\quad-\infty<\xi<\infty.
\end{equation}
Letting $a^2=x$ and substituting this equation into Eq. \eqref{eq-st12}, we obtain:
\begin{equation}\label{eq-st19a}
I_1\sqrt{\pi}
=\int_{0}^{\infty}\frac{dx}{\sqrt{x}}\,e^{-P^2x}
\int_{\mu}^{\infty}{{d\xi}\,\cfrac{\NP{\xi-\mu}^\al\NP{\xi^2+P^2}}{\NP{\xi-\mu-1}^2+x_0^2}\,e^{-\beta\xi}\,\sum_{n=0}^{\infty}\frac{(-1)^n}{2^{2n}n!}\,\frac{x^n}{(1+x)^{n+1/2}}\,H_{2n}(\xi)}.
\end{equation}
In order to justify whether it is possible to change the order of summation and integration, we should seek if the following integral converges, that is (see \cite{Titch}, pp.45; and some additional examples in \cite{Lebedev}, pp. 2, 63--64, 82 and 239),
\begin{equation}\label{eq-st19b}
\II=\int_{0}^{\infty}\EB{\frac{dx}{\sqrt{x}}\,e^{-P^2x}}
\int_{\mu}^{\infty}{{d\xi}\,\EB{\cfrac{\NP{\xi-\mu}^\al\NP{\xi^2+P^2}}{\NP{\xi-\mu-1}^2+x_0^2}\,e^{-\beta\xi}}\,\sum_{n=0}^{\infty}\EB{\frac{(-1)^n}{2^{2n}n!}\,\frac{x^n}{(1+x)^{n+1/2}}\,H_{2n}(\xi)}}.
\end{equation}
From the inequalities \eqref{ineq1}, $(1+x)^{-n/2}<1$ for $x>0$ and $\NB{H_{n}(\cdot)}\leq i^{-n}H_n\NP{i\NB{\cdot}}$ (see \cite{Lebedev}, pp. 62 and 64) we have 
\begin{equation}\label{eq-st19c}
\II\leq\EP{1+\EB{\frac{P^2+(\mu+\xi_0)^2}{\xi_0}}}\,\int_{0}^{\infty}{\frac{dx}{\sqrt{x}}\,e^{-P^2x}}
\int_{\mu}^{\infty}{{d\xi}\,\NP{\xi-\mu}^\al\,e^{-\beta\xi}\,\sum_{n=0}^{\infty}{\frac{x^n}{2^{2n}n!}\,i^{-2n}H_{2n}(i\xi)}}.
\end{equation}
From the generating function for Hermite polynomials \cite{Lebedev}:
\begin{equation}\label{eq-st19d}
e^{2wz-z^2}=\sum_{n=0}^\infty \frac{H_n(w)}{n!}\,z^n,\quad|z|<\infty,
\end{equation}
with $z=-x/4$ and $w=i\xi$, the inequality transform in the following:
\begin{equation}\label{eq-st19e}
\II\leq\EP{1+\EB{\frac{P^2+(\mu+\xi_0)^2}{\xi_0}}}\,\int_{0}^{\infty}{\frac{dx}{\sqrt{x}}\,e^{-P^2x}}
\int_{\mu}^{\infty}{{d\xi}\,\NP{\xi-\mu}^\al\,e^{-\beta\xi}\,e^{2(-x/4)(i\xi)-x^2/4}},
\end{equation}
where the integral from the r.h.s of the former inequality is convergent. Accordingly, the order of summationa and integration in Eq. \eqref{eq-st19a}  can be exchanged:
\begin{equation}\label{eq-st20}
I_1\sqrt{\pi}
=\sum_{n=0}^{\infty}\frac{(-1)^n}{2^{2n}n!}\int_{0}^{\infty}\frac{dx}{\sqrt{x}}\,\frac{x^n}{(1+x)^{n+1/2}}\,e^{-P^2x}
\int_{\mu}^{\infty}{{d\xi}\,\cfrac{\NP{\xi-\mu}^\al\NP{\xi^2+P^2}}{\NP{\xi-\mu-1}^2+x_0^2}\,e^{-\beta\xi}\,H_{2n}(\xi)}.
\end{equation}}
Here, we could decouple the original integral as an infinite sum of the product of two integrals. Needless to say that we already substituted the expression for the form factor. Now, we shall calculate the integral in $x$, and we have to consider two cases according to the values of $n$. When $n=0$, the integral is calculated from the following one \cite{Lebedev}:
\begin{equation}\label{eq-st21}
K_\nu(xz)=\sqrt{\frac{\pi}{2z}}\,\frac{x^{\nu}e^{-xz}}{\Ga\NP{\nu+\tfrac{1}{2}}}\,\int_0^\infty{e^{-xt}t^{\nu-1/2}\GP{1+\frac{t}{2z}}^{\nu-1/2}},\quad|\Arg{z}|<\pi,\Re{\nu}>-\frac{1}{2},
\end{equation}
where $K_\nu(\cdot)$ is the modified Bessel function of the second kind. Our integral is obtained if $\nu=0$, $z=\tfrac{1}{2}$ and $x=P^2$. Hence
\begin{equation}\label{eq-st22}
\int_{0}^{\infty}\frac{dx}{\sqrt{x}}\,\frac{1}{(1+x)^{1/2}}\,e^{-P^2x}
=e^{P^2/2}K_0\NP{\tfrac{1}{2}P^2}.
\end{equation}
When $n\neq0$, we reorganize the integrand of the integral in $x$ as:
\begin{equation}\label{eq-st23}
\int_{0}^{\infty}\frac{dx}{\sqrt{x}}\,\frac{x^n}{(1+x)^{n+1/2}}\,e^{-P^2x}
=\int_{0}^{\infty}\frac{dx}{x}\,\GP{\frac{x}{1+x}}^{n+1/2}\,e^{-P^2x}.
\end{equation}
Making the change of variables $y=\dfrac{x}{x+1}$:
\begin{equation}\label{eq-st24}
\int_{0}^{\infty}\frac{dx}{\sqrt{x}}\,\frac{x^n}{(1+x)^{n+1/2}}\,e^{-P^2x}
=\int_{0}^{1}dy\,y^{n-1/2}\,\frac{1}{1-y}e^{-P^2y/(1-y)}.
\end{equation}
From the generating function of Laguerre polynomials \cite{Lebedev}, i.e.,
\begin{equation}
\frac{1}{1-y}e^{-P^2y/(1-y)}=\sum_{m=0}^{\infty}L_m^0(P^2)y^m,\quad|y|<1,
\end{equation}
we find that 
\begin{equation}\label{eq-st25}
\int_{0}^{\infty}\frac{dx}{\sqrt{x}}\,\frac{x^n}{(1+x)^{n+1/2}}\,e^{-P^2x}
=\sum_{m=0}^{\infty}L_m^0(P^2)\int_{0}^{1}dy\,y^{n+m-1/2}
=\sum_{m=0}^{\infty}\frac{L_m^0(P^2)}{n+m+\tfrac{1}{2}}.
\end{equation}
Summarizing:
\begin{equation}\label{eq-st26}
\int_{0}^{\infty}\frac{dx}{\sqrt{x}}\,\frac{x^n}{(1+x)^{n+1/2}}\,e^{-P^2x}
=
\begin{cases}
e^{P^2/2}K_0\NP{\tfrac{1}{2}P^2} & n=0 \, ,\\
 & \\
\sum_{m=0}^{\infty}\dfrac{L_m^0(P^2)}{n+m+\tfrac{1}{2}} & n=1,2,3,\dotsc
\end{cases}
\end{equation}
Coming back to Eq. \eqref{eq-st20}, now we wish to calculate the integrals in $\xi$. Using Eq. \eqref{eq-st7}, we have:
\begin{multline}\label{eq-st27}
\int_{\mu}^{\infty}d\xi\,\cfrac{\NP{\xi-\mu}^\al\NP{\xi^2+P^2}}{\NP{\xi-\mu-1}^2+x_0^2}\,H_{2n}(\xi)\,e^{-\beta\xi}
=\int_{\mu}^{\infty}\,d\xi\,\NP{\xi-\mu}^\al\,H_{2n}(\xi)\,e^{-\beta\xi}
\\
-\Im{\EC{\frac{P^2-z_0^2}{x_0}\int_{\mu}^{\infty}d\xi\,\cfrac{\NP{\xi-\mu}^\al}{\xi-\mu-\xi_0}\,H_{2n}(\xi)\,e^{-\beta\xi}}}.
\end{multline}
Now, we calculate both the integrals from the above equation as follows. For the first integral, i.e.,
\[
\int_{\mu}^{\infty}\,d\xi\,\NP{\xi-\mu}^\al\,H_{2n}(\xi)\,e^{-\beta\xi}\equiv J_1,
\]
we substitute the explicit form of the Hermite polynomials 
(see \cite{Lebedev}). Hence:
\begin{align}
J_1
&=
\sum_{k=0}^{n}\frac{(-1)^k\,(2n)!}{k!(2n-2k)!}
\int_{\mu}^{\infty}\,d\xi\,\NP{\xi-\mu}^\al(2\xi)^{2n-2k}e^{-\beta\xi}
\notag\\&=
\sum_{k=0}^{n}\frac{(-1)^k\,(2n)!}{k!(2n-2k)!}\,
\int_{0}^{\infty}\,d\xi\,\xi^\al(\xi+\mu)^{2n-2k}e^{-\beta(\xi+\mu)}
\notag\\&=
{2^{2n}(2n)!}\,e^{-\beta\mu}\sum_{k=0}^{n}\frac{(-1)^k}{2^{2k}\,k!(2n-2k)!}
\sum_{s=0}^{2n-2k}\binom{2n-2k}{s}\,\mu^{2n-2k-s}\,\frac{\Ga(\al+s+1)}{\beta^{\al+s+1}}.\label{eq-st28}
\end{align}
For the latter integral, i.e., 
\[
\int_{\mu}^{\infty}d\xi\,\cfrac{\NP{\xi-\mu}^\al}{\xi-\mu-\xi_0}\,H_{2n}(\xi)\,e^{-\beta\xi}\equiv J_2,
\]
we make the change of variables $\xi\to\xi+\mu$, and next we apply the following addition theorem for Hermite polynomials, i.e.,
\begin{equation}\label{eq-st29}
H_{2n}(\xi+a)=\frac{1}{2^{2n}}\sum_{k=0}^{2n}\binom{2n}{k}H_k(\xi\sqrt{2}\,)H_{2n-k}(a\sqrt{2}\,).
\end{equation}
As a result,
\begin{align}
J_2&=e^{-\beta\mu}\int_{0}^{\infty}d\xi\,\cfrac{\xi^\al}{\xi-\xi_0}\,H_{2n}(\xi-\xi_0+\mu+\xi_0)\,e^{-\beta\xi}
\notag\\&=
\frac{e^{-\beta\mu}}{2^{2n}}\sum_{k=0}^{2n}\binom{2n}{k}H_{2n-k}\NC{(\mu+\xi_0)\sqrt{2}\,}
\int_{0}^{\infty}d\xi\,\cfrac{\xi^\al}{\xi-\xi_0}\,H_{k}\NC{(\xi-\xi_0)\sqrt{2}\,}\,e^{-\beta\xi}.\label{eq-st30}
\end{align}
Now, we use the explicit form for the Hermite polynomials, and since $z_0=\mu+\xi_0$, $J_2$ transforms as:
\begin{equation}\label{eq-st31}
J_2=\frac{1}{2^{2n}}\sum_{k=0}^{2n}\binom{2n}{k}H_{2n-k}\NP{z_0\sqrt{2}\,}
\sum_{s=0}^{[k/2]}\frac{(-1)^s\,k!\,(2\sqrt{2})^{k-2s}}{s!\,(k-2s)!}
\int_{0}^{\infty}d\xi\,\cfrac{\xi^\al}{\xi-\xi_0}\,(\xi-\xi_0)^{k-2s}\,e^{-\beta\xi}.
\end{equation}
Finally, the integral in Eq. \eqref{eq-st31} can be calculated in two parts, namely, for $k-2s=0$, we have 
\begin{equation}
\int_{0}^{\infty}d\xi\,\cfrac{\xi^\al}{\xi-\xi_0}\,e^{-\beta\xi}
=\Ga\NP{\al+1}(-\xi_0)^\al\,e^{-\beta\xi_0}\Ga\NP{-\al,-\beta\xi_0},
\end{equation}
and for $k-2s\geq1$, 
\begin{equation}
\int_{0}^{\infty}d\xi\,{\xi^\al}\,(\xi-\xi_0)^{k-2s-1}\,e^{-\beta\xi}
=\sum_{l=0}^{k-2s-1}\binom{k-2s-1}{l}(-\xi_0)^{k-2s-1-l}\,\frac{\Ga\NP{\al+l+1}}{\beta^{\al+l+1}}.
\end{equation}

\end{document}